\documentclass[pre,twocolumn,aps,showpacs]{revtex4}
\usepackage{dcolumn}
\usepackage{graphicx}
\usepackage{graphicx,amssymb,amsmath}
\usepackage{natbib}
\usepackage{bm}

\newcommand{\Nabla}{\boldsymbol{\nabla}}

\begin{document}

\title{Nanorheology of viscoelastic shells: Applications to viral capsids}

\author{Tatiana Kuriabova$^{1}$, Alex Levine$^{2,3}$}

\affiliation{$^{1}$Department of Physics and Astronomy,
University of California, Los Angeles, CA 90095\\
$^{2}$Department of Chemistry \& Biochemistry, University of
California, Los Angeles, CA 90095\\
$^{3}$California Nanosystems Institute, University of California,
Los Angeles, CA 90095}

\date{\today}

\begin{abstract}
 We study the microrheology of nanoparticle shells [Dinsmore et al.
 Science {\bf 298}, 1006 (2002)] and viral capsids
 [Ivanovska et al. PNAS {\bf 101}, 7600 (2004)] by computing the mechanical response
 function and thermal fluctuation
 spectrum of a viscoelastic spherical shell that is permeable to the
 surrounding solvent. We determine analytically the damped dynamics of
 the shear, bend, and compression modes of the shell coupled to the solvent
 both inside and outside the sphere in the zero Reynolds number limit.
 We identify fundamental length and time scales in the system,
 and compute the thermal correlation function of displacements of antipodal
 points on the sphere and the mechanical response to pinching forces
 applied at these points. We describe how such a frequency-dependent
 antipodal correlation and/or response function, which should be measurable
 in new AFM-based microrheology
 experiments, can probe the viscoelasticity of these synthetic and
 biological shells constructed of nanoparticles.
 \end{abstract}


\maketitle

\section{Introduction}
\label{sec:introduction}

Understanding the dynamics of soft, nanoporous materials will be an important area
of research both in biophysics and in the materials science
communities. This emerging
field of study is being driven by new AFM-based mechanical or rheological
measurements of supramolecular
biological materials~\cite{Dufrene:03,Radhakrishnan:04} and even entire
cells~\cite{You:99,Alonso:02}.  Perhaps the prototypical
examples of such a material are found in the plethora of viral capsids. Such shells
are constructed from a small set of proteins or their oligomers (capsomeres) arranged in an
ordered structure forming roughly a spherical shell. The packing of these
capsomeres generically leaves at least nanometer scale pores through which solvent may flow.
In addition, there is a wide variety of similarly
nanoporous synthetic structures such as linked networks of
nanoparticles~\cite{Lin:03} and colloids~\cite{Dinsmore:02,Lawrence:07}
assembled into two-dimensional shells. In both the biological and synthetic cases,
the material making up the thin shell or membrane is
constructed from identical nanoscale tiles that incorporate holes of
comparable size through which the solvent may flow.  Other examples
of porous membranes include lipid bilayers containing pore-forming
transmembrane proteins \cite{Gilbert:02,Janshoff:06} and even the large
($\sim 100$nm) cytoplasmic ribonucleoprotein vaults~\cite{Kedersha:86}, which make up a
rather ubiquitous but enigmatic component of eukaryotic cells.

\begin{figure}[htpb]
\centering
\includegraphics[width=7cm]{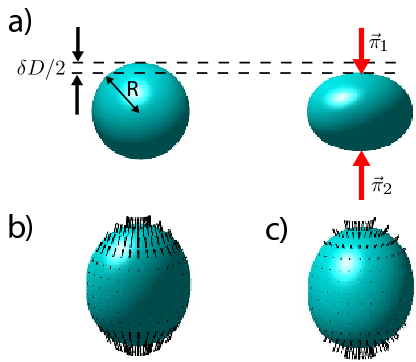}
\caption{(color online) a) A schematic illustration of an AFM-based rheological
experiment on a nanoscale shell. The (red) arrows depict the applied stresses
$\boldsymbol{\pi}_1 = -\boldsymbol{\pi}_2$
that deform the (blue) shell of radius $R$ from a sphere on the left into an oblate object
shown in the right. The total change in sphere's diameter along the line of force application is $\delta D$.
This deformation can be resolved into a linear superposition
of two sets of normal modes. In b) we show the Lennon-Brochard $\ell=2$ 
mode while in c) we show the compression dominated longitudinal sound $\ell=2$ mode. 
In each figure the (black) arrows represent the velocity field of the sphere.}
\label{fig:schematic}
\end{figure}

We focus on elucidating
the mechanical response of such a permeable shell to  a set of sinusoidally time-varying
``pinching'' forces applied at antipodal points
on the sphere -- see Fig.~\ref{fig:schematic}(a). Such forces represent the simplest characterization of
an AFM-based nanoindentation experiment performed at finite frequency.  By applying known forces at a
fixed frequency and observing the in-phase and out-of-phase response of the diameter $D$ of the spherical
shell one should be able to extract the viscoelastic properties of the shell material.
Such an experiment is the finite-frequency generalization of the
work of Michel {\em et al.}~\cite{Michel:06} and can be considered to be an active microrheological
measurement. Alternatively, one may imagine that the AFM tip can be used in
a passive manner to monitor the thermal fluctuations of this diameter.
Since a number of these physically relevant shells are
rather incompliant, their thermal fluctuations are likely to be
unresolvable. In such cases passive microrheology (i.e. the monitoring
of thermally fluctuating variables) must be supplanted by active techniques.
This distinction is of little importance for our analysis provided that the
mechanical perturbation in the active technique is small enough that
the system remains in its linear response regime. In the passive
technique one has direct access to the power spectrum of the fluctuating
quantity determined (via the fluctuation dissipation theorem) by the imaginary part
of the response function. The real part is then obtained by a Kramers-Kr\"{o}nig relation \cite{Landau:51}. In the
active measurement the real and imaginary parts of the response function are found directly from
the in-phase and out-of-phase response of the shell respectively. In this article we will compute the
full frequency-dependent, complex response function as well as the expected thermal power spectrum of
$\langle |D(\omega)|^2 \rangle$ of  each shell studied.

To understand the mechanics/dynamics of such nanoscale
objects in solution, one must consider the coupled dynamics
of soft membranes/shells and the surrounding solvent. The case of viscous shells
that are impermeable to the surrounding solvent has been well-studied in the context of
microemulsions and vesicles~\cite{Milner:87}. The focus of our current study, however, is the dynamics
of membranes or shells that are permeable to the surrounding solvent on time-scales
relevant to the experiment. The effect of solvent permeation, which we quantify below, is to
create a new dissipative stress on the shell related to solvent permeation.
These dissipative (out-of-phase) stresses associated with fluid permeation must also be taken into account
for the quantitatively correct interpretation of such measurements as described above. In particular
in the case of a membrane or shell that has an inherent viscoelastic response, one may ask whether
it remains possible to distinguish the dissipative stresses due to from the material from those associated
with fluid permeation.

In this article we quantify and explore the effects of membrane permeability on
the dynamics of both flat and spherical membranes. We parameterize the continuum mechanics of
the shells by a permeation coefficient, a bending modulus, an area compression modulus,
and a two-dimensional, in-plane shear modulus. These latter two moduli we will allow to be
viscoelastic, i.e. complex and frequency dependent. We discuss the interpretation of this
generalization below.  We show that solvent permeability leads to a new dissipative stress in the system
that changes the relaxation rates of each deformation mode of
the system.  We show also that, on a spherical shell, these normal modes are superpositions of
compression and bending deformations.  We then consider four test cases to explore the
interrelated roles of shell porosity and viscoelasticity in determining their mechanical response.

First, we examine the expected response of a purely elastic, porous shell to 
isolate the dissipative effects of the interaction
of the porous shell with the surrounding solvent.
We then consider the case of a highly incompressible, but purely
viscous membrane of varying porosity. Such systems are reminiscent of giant
unilamellar vesicles (GUVs)~\cite{Dietrich:01,Veatch:02,Veatch:03}
containing pore-forming proteins~\cite{Ojcius:91,Abrami:98,Kaneko:04}.

Finally, we consider two simple models of a viscoelastic shell inspired by viral capsids. In the first
case we assume that the shell has a purely elastic response to compression but a viscoelastic response
to in-plane shear stresses. Such a viscoelastic response to in-plane shear may result from plastic rearrangements
of capsomere proteins under applied stresses. Thus, the proteinaceous shell of a virus may deform like an
amorphous solid. In the second case we assume an elastic response to in-plane shear stress, but the
dissipative relaxation of in-plane compressional stress. Such stress relaxation should occur in cases where the
capsomeres have internal degrees of freedom associated with allosteric
conformational transitions \cite{Conway:95, Steven:05,Guerin:07} that change their cross-sectional area or
packing density. To model this effect we assume a compressional stress relaxation time on a scale typical of
known allosteric transitions in proteins.  In both cases we employ the simplest model of viscoelasticity, the
Maxwell model~\cite{Bird:77}, which has a single stress-relaxation time scale.

The central question we address
is whether it is possible to extract the dissipative mechanics of the capsid proteins
from the proposed AFM-based microrheological experiment. It is possible that the dissipative
stresses associated with internal rearrangements of the capsomeres or plastic deformations of the shell
will be lost against the background of dissipative forces associated with solvent flow through
the porous membrane. In both cases, owing to a separation of time scales, the nanoindentation experiment should
be able to resolve the expected viscoelastic response of the shell. We show, however,
that in the former case of a viscoelastic shear response, the mechanical signature
is quite subtle in the real part of the response function due to the dominance of the bending and
compression moduli in the response function. However, the imaginary or out-of-phase part of the response function
is capable of recording this effect. In the latter case
the effect of compressional stress relaxation on time scales typical of protein conformational change
should give a dramatic mechanical signature in the imaginary part of the response function
so that finite-frequency nanoindentation studies should be an excellent
probe of these complex mechanics.

The remainder of the paper is organized as follows. In section~\ref{sub:plane-membrane} we calculate the effect of
porosity on the undulation dynamics of a flat membrane in a viscous solvent. We also discuss expected values of the
permeation coefficient for a variety of physically important systems. The calculation of these coefficients is
discussed in appendix~\ref{Appendix-A}. In section~\ref{sub:shell-mechanics}
we consider the deformations of a spherical membrane where the curvature of the undeformed structure couples the
previously studied bending mode to in-plane compression. To incorporate the mechanical effects of
solvent hydrodynamics in the case of spherical shells, we recount and apply the pioneering
work of Lamb~\cite{Lamb:32} and
Brenner~\cite{Brenner:64} on low-Reynolds number flows inside and outside a sphere in
section~\ref{sub:hydrodynamics}. Using the results of the previous calculations we plot and discuss the decay rates
of the linearly independent modes of the shell in section~\ref{sub:modes}.  
In section~\ref{sec:response-function}
we use the dynamics of these linearly-independent modes to construct the 
complex, frequency-dependent response
function of the shell to antipodal pinching forces. 
In sections~\ref{sub:results-elastic}, \ref{sub:results-viscous}, and
\ref{sub:results-viscoelastic} we plot the response functions and
predicted thermal power spectra of elastic, viscous, and the two 
classes of viscoelastic shells introduced above.
Finally, in section~\ref{sec:discussion}, we compare 
these results and discuss the implications of our work for future
nanoindentation-based rheology measurements.

\section{Calculation}
\label{sec:calculation}

\subsection{The plane permeable membrane}
\label{sub:plane-membrane}

The overarching feature of these systems is that fluid flow through
the shell or membrane (permeation) can occur on timescales
comparable to the relaxation of surface deformations. Thus, the
velocity of the fluid at the membrane surface in the direction along
the local normal of that surface will not be equal to the normal
velocity of the membrane.  In the limit of high membrane porosity,
the fluid will essentially pass though the membrane undisturbed.
Consequently the hydrodynamic interaction of the membrane with
itself, which is known to qualitatively alter the spectrum of the
decay rates of membrane undulations~\cite{Brochard:75} is suppressed
and replaced by dynamics consistent with local drag. As the porosity
or permeability of the membrane is reduced, the system continuously
recovers these well-known hydrodynamic self-interactions that
control the structure of the spectrum of undulatory decay rates. 
To better understand this feature we address the
dynamics of a {\em plane} porous elastic membrane in a viscous
Newtonian solvent. We assume that the dynamics are all occurring at
low Reynolds number. For a flat membrane or interface the normal 
modes of the coupled interface and solvent system involve bending 
in-plane shear and compression~\cite{Levine:02}. Since we will 
concentrate on indentation-based measurements, we consider 
only the undulation mode of the flat membrane.

A zero-tension membrane is characterized by a single
bending modulus $\kappa$ so that free energy of the membrane may be
written as
\begin{equation}
\label{bending-flat}
{\cal F}=\frac{\kappa}{2}\int \left[ \nabla_\perp^2 w
\right]^2dxdy,
\end{equation}
in terms of normal displacement $w$. Here $\nabla_\perp$ is the
in-plane gradient operator acting on membrane coordinates $x,y$. In
the above we implicitly assumed small deformations of the surface
from its flat equilibrium shape. The restoring stresses acting on
the deformed membrane are given by
\begin{equation}
\label{elforce} f=-\frac{\delta {\cal F}}{\delta w}=-\kappa
\nabla_\perp^4 w.
\end{equation}
The equation of motion for the fluid of viscosity $\eta$ in the 
limit of vanishing Reynolds
number is the linearized Navier-Stokes equation for an
incompressible fluid:
\begin{eqnarray}
&\Nabla\cdot{\bf v}=0& \\
&\displaystyle \rho \frac{\partial {\bf v}}{\partial t}=\eta
\nabla^2 {\bf v} - \Nabla p,
\end{eqnarray}
where $p$ is the pressure.
The interaction between the fluid and the membrane is governed by
three boundary conditions enforced at the membrane lying in the $xy$
plane. First, we require Darcy's law to be obeyed so that the flux
of the fluid through a unit area of the membrane is proportional to
the pressure difference between the two sides \cite{Batchelor:67}. Thus
\begin{equation}
\label{Darcy}
 \alpha\left(v_z\Big{|}_{z=0} - \frac{dw}{dt} \right)=p^- -  p^+ ,
\end{equation}
where $p^{+,-}$ are the pressures above ($z>0$) and below ($z<0$)
the membrane respectively. We discuss the estimation of the Darcy
coefficient $\alpha$ in terms of the microscale structure of the
membrane in Appendix \ref{Appendix-A}. Expected values of the
permeation coefficient for a variety of systems are given in Table 1.
This first boundary condition relates the normal velocities of the
membrane and fluid at the membrane surface by equating the normal
stress difference across the membrane to the fluid flow through it.
In addition we must insist upon the continuity of the normal
component of the fluid velocity so that
\begin{equation}
v_z^+(0)=v_z^-(0),
\end{equation}
where $v_z^{\pm}(0) = \lim_{z\to 0^\pm} v_z(z)$. For the
tangential velocities, we enforce a no-slip boundary condition on
the fluid at the upper and lower boundaries of the membrane:
\begin{equation}
v_x^+(0)=v_x^-(0)=0.
\end{equation}
Finally, we require the fluid velocity field and pressure to vanish
infinitely far from the membrane so that
\begin{equation}
{\bf v}(\pm \infty)=0,\quad p(\pm \infty)=0.
\end{equation}

Ignoring the inertia of the membrane, we require force balance
between the elastic restoring stress $f$ from Eq.~(\ref{elforce}),
the viscous stresses $\sigma^+_{zz}$ and $\sigma^-_{zz}$ exerted by
the fluid on both sides of the membrane and an externally imposed
stress, $\pi_e$ so that
\begin{equation}
(\sigma_{zz}^- -\sigma_{zz}^+)\Bigl{|}_{z=0} +\, \pi_e
=\kappa\Nabla^4 w.
\end{equation}
\begin{table}
\caption{\label{table}Dimensionless permeation coefficients (Eq.~(\ref{bar-alpha}))
for nanoscale capsules}
\begin{ruledtabular}
\begin{tabular}{ccc}
System & Permeation coefficient & Ref\\\\
Rotavirus & 50 & \cite{Johnson:97} \\
S-layer shell& $10^4 $& \cite{S-layer:97}\\
Colloidosome& $10^4-10^5$ & \cite{Dinsmore:02}\\
GUV& $10-\infty$ & \cite{Gilbert:02,Janshoff:06}\\
\end{tabular}
\end{ruledtabular}
\end{table}
To compute the decay rate of membrane undulations having wave vector
$q$, we consider a sinusoidally applied normal stress $\pi_e$
\begin{equation}
\pi_e=\pi\,e^{iqx-i\omega t},
\end{equation}
and look for solutions of the membrane deformation taking form of
\begin{equation}
w = w_0 e^{iqx-i\omega t}.
\end{equation}
Details of the calculation are presented in Appendix
\ref{Appendix-B}. In the limit that viscous stresses dominate
inertial ones in the fluid, {\em i.e.} $\omega\rho\ll \eta q^2$, we
find that the amplitude of the driven undulations is given by
\begin{equation}
\label{flat-ans} w_0=-\frac{ \pi\, \left[ \frac{\alpha
}{2\,q\,\eta}+2 \right]} {2\,i\,\omega\,\alpha -\kappa q^4 \, \left[
\frac{\alpha }{2\,q\,\eta}+ 2\right]}.
\end{equation}
The poles of the response function shown in Eq.~(\ref{flat-ans}),
determine the relaxation rate of the freely decaying surface. We
find that the decay rate takes the form
\begin{equation}
\label{decay-rate}
\omega=-i \,\kappa\,\frac{q^3}{4\eta} - i\,
\frac{\kappa q^4}{\alpha}.
\end{equation}

In the limit of an impermeable membrane, i.e. $\alpha\to
\infty$, we recover the Lennon Brochard result~\cite{Brochard:75}.
As $\alpha$ becomes smaller reflecting the increasing permeability
of the membrane, the decay rate that scales as $q^3$ crosses over to
$q^4$ scaling. This small-$\alpha$ dynamics is consistent with a
``free-draining'' assumption that the fluid merely exerts a local
drag on the membrane, but does not contribute to a long-ranged
hydrodynamic interaction between distant parts of the surface.  For
any finite permeability, one may note from Eq.~(\ref{decay-rate}) that
there is a cross-over length $l \sim \eta/\alpha$ above which the
decay of surface undulations is controlled by long-range
hydrodynamic interactions, while undulations of a wavelength less
than $l$ experience the free-draining dynamics.

Given this intuition gained from the more simple question of small
undulations of a planar, permeable membrane, we now turn to the
problem of a spherical shell that is a more appropriate model of a
viral capsid or its synthetic mimics.

\subsection{Mechanics of a porous spherical shell.}
\label{sub:shell-mechanics}

Extending our analysis of the dynamics of a plane porous membrane to the
case of spherical ones involves two separate complications. The
first is due to the spherical geometry of the undeformed membrane.
In the case of a flat interface the in-plane shear and compression
modes of the material decouple from the out-of-plane bending modes
to linear order in the bending deformation~\cite{Landau:59}. This is
not the case for the sphere.  For small deformations of surfaces
having finite curvature, out-of-plane deformation is coupled at first
order to in-plane dilatation and shear~\cite{Kroll:93}.

We write the elastic bending free energy ${\cal F}_b$ of the
spherical shell using the well-known Helfrich
form~\cite{Helfrich:73,Milner:87,Kroll:93}
\begin{equation}
\label{bending-Helf}
{\cal F}_b=\frac{\kappa}{2}\int d^2s (K^{\alpha}_{\alpha}
-c_0)^2,
\end{equation}
where $d^2s$ is an element of area on the surface, $\kappa$ is once
again the bending rigidity, $K^{\alpha}_{\alpha}$ is the trace of
the curvature tensor, and $c_0$ is the spontaneous curvature. The 
Greek indices run over the (angular) coordinates of the undeformed sphere. The
curvature tensor on the spherical membrane of radius $R$ may be
written in terms of the out-of-plane deformation, $w$,  as
\begin{equation}
\label{curavature} K^{\alpha}_{\beta}=D^{\alpha}D_{\beta} w + \delta^{\alpha}_{\beta}\,w/R^2+
\delta^{\alpha}_{\beta}/R.
\end{equation}

In order to address out-of-plane deformation on the sphere, we
must include the in-plane elastic stresses associated with the
dilatation and shear of the membrane.  To do this we introduce a
two-dimensional (2D) covariant strain tensor given by
\begin{equation}
\label{strain-tensor}
E^{\alpha}_{\beta}=\frac{1}{2} (D^{\alpha}
t_{\beta} +D_{\beta} t^{\alpha}) +\delta^{\alpha}_{\beta}\,w/R,
\end{equation}
where $\boldsymbol{t}$ is a displacement vector in tangent space of the
sphere. Using this decomposition of the
deformation into in-plane and out-of-plane motion, the most general
displacement of material elements of sphere can be expressed in the
form
\begin{equation}
\label{def-varepsilon}
\boldsymbol{\varepsilon}=w\, \hat{\bf r}+{\bf
t}.
\end{equation}
The in-plane elastic energy of the sphere can be written as a
surface integral of a scalar elastic energy density
\begin{equation}
\label{strain} {\cal F}_e=\int d^2s \left[\mu
E^{\beta}_{\alpha}E^{\alpha}_{\beta} +\frac{\lambda}{2}
(E^{\alpha}_{\alpha})^2\right],
\end{equation}
where $\mu$ and $\lambda$ are the two  2D Lam\'{e} coefficients
required to describe the elasticity of isotropic materials. It is
not clear {\em a priori} that the isotropic elasticity is sufficient
to accurately describe the mechanics of a viral capsid or other
ordered structures, but this choice introduces the minimal set of
unknown parameters and greatly simplifies the analysis. In
discussions of the equilibrium spectrum of shape deformations of
these membranes we will consider the viscoelastic generalization of
Eq.~(\ref{strain}) where the Lam\'{e} coefficients are complex
functions of frequency so that the strain energy involves an
integral over the history of prior deformations.

Combining Eqs.~(\ref{strain-tensor}) and (\ref{strain}) it is clear
that the strain energy contains a term bilinear in out-of-plane and
in-plane deformation. This shows that the curvature of the surface
does indeed induce a linear order coupling of the in-plane
deformation to forces along the local membrane normal. The total
elastic free-energy $\cal{F}_{\rm el}$ of membrane is then expressed as
the sum of Eqs.~(\ref{bending-Helf}) and (\ref{strain}), i.e. it is given
by ${\cal F}_{\rm el}={\cal F}_b +{\cal F}_e$.

We now look for the normal modes of the deformation of the sphere.
Based on intuition from flat space, we write the in-plane
displacement field ${\bf t}$ as the sum of an irrotational and a
solenoidal part
\begin{equation}
\label{t-decomp}
t_{\beta}=D_{\beta}\Psi +
\gamma_{\alpha\beta}D^{\alpha}\chi,
\end{equation}
where $\Psi$ and $\chi$ are two scalar fields defined on the surface
of the sphere and $\gamma_{\alpha\beta}$ is the alternating tensor.
From Eq.~(\ref{t-decomp}) one notes that $\Psi$ determines the
dilatational deformation of the system, while $\chi$ describes the
density-preserving shear modes of the membrane.

Using this decomposition, the elastic free energy of the spherical
shell is given by
${\cal F}_{\rm el}={\cal F}_1[w,\Psi]+{\cal F}_2[\chi]$, where
\begin{eqnarray}
\lefteqn{{\cal F}_1[w,\Psi]=\int d^2s \Biggl\{ \frac{\kappa}{2}
w(\boldsymbol{\Delta}_{\perp}+2/R^2)^2 w \nonumber}\\
&& { }+\frac{2\kappa}{R^2}\,(2/R-c_0)\,w + \frac{2 K}{R^2}\, w^2  + \frac{2 K}{R}\, w\,
\boldsymbol{\Delta}_{\perp}\Psi\nonumber \\
&&\label{decomp-inplane-elastic}
 { } + \frac{1}{2}(K+\mu)(\boldsymbol{\Delta}_{\perp}\Psi)^2
+\frac{\mu}{R^2}\Psi (\boldsymbol{\Delta}_{\perp}\Psi)
\Biggr \}
\end{eqnarray}
and
\begin{eqnarray}
{\cal F}_2[\chi]=\int d^2s\left\{\frac{\mu}{2} \boldsymbol{\Delta}_{\perp}\chi
\left[\boldsymbol{\Delta}_{\perp}\chi+2\chi/R^2 \right]\right \};
\end{eqnarray}
$\boldsymbol{\Delta}_{\perp}$ is the two-dimensional in-plane
Laplacian and $K=\mu+\lambda$~\cite{Kroll:93}. Since there is no coupling between
$\left\{ w,\Psi \right\}$ and $\chi$, purely radial deformations of
the sphere will excite only the compression modes related to $\Psi$.
Therefore, in the interest of studying microrheological approaches
to the measurement of the membrane mechanics via nanoindentation
studies, we may neglect ${\cal F}_2[\chi]$. Of course, to address
the question of the dynamics of the porous, spherical membrane we
must consider the (visco-)elastic object described above coupled to
the solvent flows around and through it. In order to do so, we
examine the fluid motions inside and outside the sphere in the
zero Reynolds number or creeping flow regime.

\subsection{Hydrodynamics of the spherical shell}
\label{sub:hydrodynamics}

It has long been recognized that studying this hydrodynamics of a
fluid confined to either the interior or exterior of a sphere is
facilitated by recasting the Stokes equation in spherical
coordinates and expanding the solutions in a basis of the
appropriate solid spherical
harmonics~\cite{Lamb:32,Brenner:64,Happel:83}. These results were
used~\cite{Schneider:84} to explore the dynamics of an impermeable
elastic shell immersed in a viscous fluid.  In the interests of
presenting a self-contained analysis, we recapitulate some of the
earlier work on the hydrodynamics of the problem. We then discuss
and expand upon the work of Schneider {\em et
al.}~\cite{Schneider:84, Jenkins:77} to introduce permeability, and to correct
the mechanical coupling between bending and in-plane dilatation that
is imposed by the curvature of the membrane as demonstrated in
Eq.~(\ref{strain-tensor}).

To set the boundary conditions in a manner similar to that discussed
for the case of an undulating permeable plane, we enforce the
continuity of the radial components of the fluid velocity normal to
the surface of the membrane so that
\begin{equation}
\label{v-radial} \left. v_r^{\rm in} \right|_{r=R} =\left.
v^{\rm out}_r\right|_{r=R},
\end{equation}
where here and in the following we label the field associated with 
the fluid in the interior by ``in'' and on the exterior of the spherical shell by ``out.''
We also require a no-slip boundary condition on the components of
the velocity field in the tangent plane of the membrane so that
\begin{eqnarray}
\label{no-slip}
{\bf v}^{\rm in}_{\perp} = {\bf v}^{\rm out}_{\perp}= {\bf V}_{\perp},
\end{eqnarray}
where the velocity of the membrane ${\bf V}$ is given by
\begin{equation}
\label{V} {\bf
V}=\frac{d\boldsymbol{\varepsilon}}{dt}=\frac{dw}{dt}\hat{\boldsymbol{
r}}+\boldsymbol{\nabla}_{\perp} \frac{\partial\psi}{\partial t}
\end{equation}
using Eq.~(\ref{def-varepsilon}). Here $\boldsymbol{\nabla}_{\perp}$
is the gradient operator in the tangent plane of the sphere and
$\hat{\boldsymbol{r}}$ is the local outward unit normal.

To couple the flow of the fluid through the membrane to the
deformation of it, we insist on force balance at the membrane so
that
\begin{equation}
\label{force-balance}
{\bf F}={\bf F}_{\rm fluid},
\end{equation}
where ${\bf F}$ is the viscoelastic restoring force acting on the
membrane due to its deformation history, and ${\bf F}_{\rm fluid}$
is the negative of the difference between the viscous stresses
$\sigma_{ij}^{\rm in,out}$ on the membrane due to fluid flow inside
and outside of it respectively,
\begin{equation}
\label{force-fluid}
({\bf F}_{\rm fluid})_j = \hat{r}_k \left(
\sigma^{\rm in}_{kj}- \sigma^{\rm out}_{kj}\right).
\end{equation}
Finally, the local difference in the normal velocity of the fluid
and the membrane is governed by Darcy's law
\begin{equation}
\label{flux} \alpha \left[v_r \Big |_{r=R} -\frac{dw}{dt}\right]=(\sigma^{\rm in}_{rr} -\sigma^{\rm out}_{rr}),
\end{equation}
where $\alpha$ is the permeation parameter introduced in Eq.~(\ref{Darcy}).

The rotational symmetry of the problem ensures that the coupled
membrane deformations and fluid motion corresponding to each
spherical harmonic decouple from all other such terms. We can expand
the fluid velocity field inside and outside the spherical membrane
in terms of two scalar potentials,$\Phi, \chi^{\rm
f}$ and pressure $p$~\cite{Lamb:32,Brenner:64,Happel:83}. Due to incompressibility
and the Stokes equation, one can show that these functions and the
pressure field $p$ each satisfy a Laplace equation,
\begin{equation}
\label{Laplace} \Nabla^2 p=0, \quad \Nabla^2 \Phi =0, \quad
\Nabla^2\chi^{\rm f} =0.
\end{equation}
It can be shown that the field $\chi^{\rm f}$ is the source of fluid
vorticity. To analyze the effects of indentation experiments on the
sphere, we will study membrane deformation without in-plane shear.
The remaining deformations do not couple to fluid vorticity so we
may set $\chi^{\rm f}=0$ hereafter. We expand the two remaining
functions $p$, $\Phi$ in terms of solid spherical harmonics in the
interior and exterior of the spherical membrane:
\begin{eqnarray}
&p^{\rm  in}=&\sum_{\ell m }P^{\rm in}_{\ell m}(t)\,r^\ell \,Y_{\ell m}(\theta,\phi)
\label{expansion1} \\
&\Phi^{\rm in} =&\sum_{\ell m} \Phi_{\ell m}^{\rm in}(t)
r^\ell  \,Y_{\ell m}(\theta,\phi)\label{expansion2} \\
&p^{ \rm out}=&\sum_{\ell m}P^{\rm out}_{\ell m}(t)
r^{-\ell -1}\, Y_{\ell m}(\theta,\phi)\label{expansion3} \\
&\Phi^{\rm out}=&\sum_{\ell m} \Phi_{\ell m}^{\rm out}(t) r^{-\ell
-1}\,Y_{\ell m}(\theta,\phi)\label{expansion4}.
\end{eqnarray}
In terms of the basis of solid spherical harmonics we may write the
velocity field in the exterior and interior of the sphere as a sum
over angular modes $n=(\ell,m)$
\begin{eqnarray}
\label{v-field-sphere}
\lefteqn{{\bf v}=\sum_{n=-\infty}^{\infty}
\left[\Nabla \times ({\bf r}\, \chi^{\rm f}_n) + \Nabla \Phi_n \right.} \\
&&\left. { } +\frac{(n+3)}{2\eta(n+1)(2n+3)}\, r^2 \,\Nabla p_n -
\frac{n}{\eta(n+1)(2n+3)}\, {\bf r}\, p_n \right]. \nonumber
\end{eqnarray}
Using Eqs.~(\ref{v-field-sphere}) and
(\ref{expansion1})-(\ref{expansion4}) we find that the continuity of
the radial fluid velocity implies
\begin{eqnarray}
\frac{\ell\,R^{\ell+1}}{2\eta (2\ell+3)}
P^{\mbox{\footnotesize in}}_{\ell m} +
\frac{\ell}{R}\,R^{\ell}\Phi^{\mbox{\footnotesize in}}_{\ell m}
+\frac{(\ell +1)R}{2\eta(1-2\ell)}\,
\frac{P^{\mbox{\footnotesize out}}_{\ell m}}{R^{\ell+1}} \nonumber\\
{}+\frac{(\ell +1)}{R}\,
\frac{\Phi^{\mbox{\footnotesize out}}_{\ell m}}{R^{\ell +1}}=0.
\label{one}
\end{eqnarray}
The tangential velocity of the fluid in the absence of vorticity, on
the other hand, is given by
\begin{eqnarray}
\label{v-t-in}
\lefteqn{{\bf v}^{\mbox{\footnotesize in}}_{\perp} =}\\
&&\nonumber \boldsymbol{\nabla}_{\perp} \sum_{\ell m}
\left[\Phi_{\ell m}^{\mbox{\footnotesize in}}(t)+
\frac{(\ell+3)r^2}{2\eta (\ell+1)(2\ell+3)}\,
P^{\mbox{\footnotesize in}}_{\ell m}(t)\right]r^\ell
\end{eqnarray}
and
\begin{eqnarray}
\label{v-t-out}
\lefteqn{{\bf v}^{\mbox{\footnotesize out}}_{\perp} =}\\
&& \nonumber \boldsymbol {\nabla}_{\perp}\sum_{\ell m}
\left[\Phi_{\ell m}^{\mbox{\footnotesize out}}(t)
-\frac{(2-\ell)r^2}{2\eta \ell(1-2\ell)} \,P_{\ell m}^{\mbox{\footnotesize out}}(t)
\right]r^{-\ell -1},
\end{eqnarray}
where we have used Eqs.~(\ref{expansion1})-(\ref{v-field-sphere}).

We now expand the membrane deformations in spherical harmonics in a
manner analogous to our expansion of the fluid velocity field by
writing
\begin{eqnarray}
\label{w-expansion}
&w(\theta,\phi,t)=&\sum_{\ell m}w_{\ell m}(t) Y_{\ell m}(\theta,\phi)\\
&\Psi(\theta,\phi,t)=&\sum_{\ell m}\Psi_{\ell m}(t) Y_{\ell
m}(\theta,\phi).
\label{psi-expansion}
\end{eqnarray}
Using Eqs.~(\ref{w-expansion}) and (\ref{psi-expansion}) along with
Eqs.~(\ref{v-t-in}) and (\ref{v-t-out}) in Eq.~(\ref{no-slip}), we
find that the no-slip condition on the tangential velocity field at
the membrane gives two equations
\begin{eqnarray}
R^{\ell}\Phi^{\mbox{\footnotesize in}}_{\ell m}+\frac{(\ell +3)
R^{\ell+2}}{2\eta(\ell +1)(2\ell +3)}P^{\mbox{\footnotesize in}}_{\ell m}
- \dot{\Psi}_{\ell m}=0
\label{two}
\end{eqnarray}
and
\begin{eqnarray}
\frac{\Phi^{\mbox{\footnotesize out}}_{\ell m}}{R^{\ell+1}}-
\frac{(2-\ell)R^2}{2\eta \ell (1-2\ell )}
\frac{P^{\mbox{\footnotesize out}}_{\ell m}}{R^{\ell+1}} -
 \dot{\Psi}_{\ell m}=0,
\label{three}
\end{eqnarray}
where the $\dot{ }$ denotes the time derivative.

We have now accounted for the kinematic boundary conditions on the
fluid and membrane velocity at the surface of the sphere. We now
require the balance of radial hydrodynamic and viscoelastic
stresses at this surface to satisfy Darcy's law. To ensure force
balance at the membrane we demand the equality of the vectors in
Eq.~(\ref{force-balance}). Clearly this results in three scalar
equations, but, it is computationally convenient to chose those
scalars in a rather nontransparent manner~\cite{Brenner:64}.
We first set the components of the forces normal to the
membrane equal to each other:
\begin{equation}
\label{F-normal}
\hat{\bf r}\cdot{\bf F}_{\mbox{\footnotesize fluid}}=\hat{\bf r}\cdot{\bf F}.
\end{equation}
We then  set the divergence of the two force fields equal to each
other at the surface of the membrane using the vector identity
\begin{equation}
\label{F-identity}
\left[({\bf r}\cdot\Nabla)\left(\frac{\bf r}{r}\cdot
{\bf F}_{\mbox{\footnotesize fluid}}\right )
-r\,\Nabla\cdot{\bf F}_{\mbox{\footnotesize fluid}}\right]_{r=R}=-R\,
\Nabla \cdot{\bf F}.
\end{equation}
Finally we also set the curls of these vector fields equal to each
other on the surface of the membrane,
\begin{equation}
\left[{\bf r}\cdot \Nabla
\times {\bf F}_{\mbox{\footnotesize fluid}} \right]_{r=R}= {\bf r}\cdot
\Nabla \times {\bf F}.
\label{curls}
\end{equation}
These curls, in fact, vanish since we have excluded in-plane shear
deformations of the membrane and vorticity of the fluid so this
additional boundary condition is automatically satisfied.

We now enforce Eq.~(\ref{F-normal}). The radial component of the
forces on the membrane due to the fluid~\cite{Brenner:64} is given
by
\begin{eqnarray}
\lefteqn{\hat{\bf r}\cdot{\bf F}_{\rm fluid}=\eta\sum_{\ell m}\left[ \frac{-2(\ell +1)(\ell +2)}{R^2}\,
\frac{\Phi_{\ell m}^{\rm out}(t)}{R^{\ell+1} }
\right. }\nonumber\\
&&{ }+\frac{\ell^2+3\ell -1}{\eta\,(2\ell -1)}
\,\frac{P_{\ell m}^{\rm out}(t)}{R^{\ell+1}}+
\frac{2\ell (\ell -1)}{R^2}\,R^{\ell}\Phi_{\ell m}^{\rm in}(t)\nonumber\\
&&\left. { }+\frac{\ell^2-\ell -3}{\eta (2\ell +3)}R^{\ell}P^{\rm in}_{\ell m}(t)\right ]
Y_{\ell m}(\theta,\phi).\label{F-n-expanded}
\end{eqnarray}
The normal component of the viscoelastic forces on the membrane can
be written in terms of $w$ and $\Psi$, the radial and in-plane
longitudinal deformations of the membrane. These forces are computed
by a derivative of ${\cal F}_1$:

\begin{eqnarray}
\label{F-n-capsid}
\lefteqn{{\bf F}\cdot \hat{\bf r}=-\frac{\delta {\cal F}_1}{\delta w}\,
\hat{\bf r}=\hat{\bf r}\left[ -\frac{2\,\kappa}{R^2}\, (2/R-c_0)-
\frac{4 K}{R^2}\,w \right.\nonumber}\\
&&\left.{} -\kappa\left(\boldsymbol{\Delta}_{\perp}+2/R^2\right)^2 w
-\frac{2 K}{R}{\boldsymbol{\Delta}_{\perp}}\Psi\right].
\end{eqnarray}
For the case of a viscoelastic membrane Eq.~(\ref{F-n-capsid})
involves an integral over the deformation history of the membrane.
In the frequency domain, however, we incorporate the viscoelastic
case by letting the elastic moduli become frequency-dependent
complex quantities. The same applies to the Eq.~(\ref{nabla-F})
below. The spontaneous curvature $c_0$ appearing in the force law
Eq.~(\ref{F-n-capsid}) generates an additional,
deformation-independent normal force on the membrane that is
compensated by internal stresses in the membrane. Since we do not
consider the nonlinear response of the elastic material such
internal stresses do not affect the dynamics. Consequently, we set
$2/R-c_0=0$ in the remainder of this work.

Expanding Eq.~(\ref{F-n-capsid}) in spherical harmonics and equating
these normal viscoelastic forces to the hydrodynamic ones from
Eq.~(\ref{F-n-expanded}) as required by Eq.~(\ref{F-normal}) leads
to
\begin{eqnarray}
&&\nonumber \displaystyle{}-2\,\frac{K}{R^3}\,\ell (\ell +1)\Psi_{\ell m}\\
&&{}+\Bigl(\frac{\kappa}{R^4} (\ell +2)^2(\ell -1)^2+4\frac{K}{R^2}\Bigr)\,
w_{\ell m}\nonumber\\
&&{}-\frac{\ell^2+3\ell -1}{(1-2\ell)}\,
\frac{P_{\ell m}^{\rm out}}{R^{\ell+1}} -\frac{2\eta(\ell +1)(\ell +2)}{R^2}\,
\frac{\Phi_{\ell m}^{\rm out}}{R^{\ell+1}}\nonumber \\
&&{}+\frac{\ell^2-\ell -3}{(2\ell +3)}R^{\ell}P^{\rm in}_{\ell m}
+2\eta \frac{\ell (\ell -1)}{R^2}R^{\ell}\Phi_{\ell m}^{\rm in}=0.
\label{normal-expanded}
\end{eqnarray}

Using the same analysis to demand force balance for the tangential
components of the forces, we determine LHS of Eq.~(\ref{F-identity})
to be
\begin{eqnarray}
\lefteqn{\left[({\bf r}\cdot\Nabla)\left(\frac{\bf r}{r}\cdot {\bf F}_{\rm fluid}\right )
-r\,\Nabla\cdot{\bf F}_{\rm fluid}\right]_{r=R}=\nonumber}\\
&&\eta\sum_{\ell m}\left[\frac{2 (\ell +1)(\ell +2)^2}{R^2}\,
\frac{\Phi_{\ell m}^{\rm out}(t)}{R^{\ell+1}}\right.\nonumber\\
&&{}+ \frac{2\ell (\ell -1)^2}{R^2}\,R^{\ell}\Phi_{\ell m}^{\rm in}(t)\nonumber\\
&&{}+\frac{\ell^3+3\ell^2+5\ell  -3}{\eta (1-2\ell)}\,
\frac{P_{\ell m}^{\rm out}(t)}{R^{\ell+1 }}\nonumber\\
&&{}+\left.\frac{\ell^3+2\ell +6}{\eta(2\ell +3)}\,R^{\ell}P^{\rm in}_{\ell m}(t)
\right]Y_{\ell m}(\theta,\phi).
\label{F-identity-expanded}
\end{eqnarray}
The RHS of Eq.~(\ref{F-identity-expanded}) can be written in terms
of the deformation field of the membrane,
\begin{eqnarray}
\label{F-t-capsid}
\lefteqn{\hat{\bf r}\times(\hat{\bf r}\times{\bf F})-\frac{\delta {\cal F }_1}{\delta \Nabla_{\perp} \Psi}=\nonumber} \\
&&\Nabla_{\perp}\left(\frac{2K}{R} w + (K+\mu)\Nabla_{\perp}^2\Psi +
\frac{2\mu}{R^2}\,\Psi   \right)
\end{eqnarray}
and
\begin{eqnarray}
\nonumber
&&-R\,\Nabla \cdot{\bf F}=\frac{8 K}{R^2}+ 2\kappa\left(\Nabla_{\perp}^2+
\frac{2}{R^2}\right)^2 w - 2K \Nabla_{\perp}^2 w\\
&&{ }-\,R(K+\mu)\Nabla_{\perp}^4\Psi-\frac{2\mu}{R}\Nabla_{\perp}^2\Psi
+\frac{4K}{R}\Nabla_{\perp}^2\Psi.
\label{nabla-F}
\end{eqnarray}
Expanding Eq.~(\ref{nabla-F}) in spherical harmonics and equating it
to Eq.~(\ref{F-identity-expanded}) we find
\begin{eqnarray}
&&\frac{\ell^3+2\ell +6}{(2\ell +3)}\,R^{\ell}P^{\rm in}_{\ell m}
+ \frac{2\,\eta\, \ell (\ell -1)^2}{R^2}\,R^{\ell}\Phi_{\ell m}^{\rm in}\nonumber \\
&&{}+\frac{\ell^3+3\ell^2+5\ell-3}{(1-2\ell)}\,\frac{P_{\ell m}^{\rm out}}{R^{\ell+1}}
+ \frac{2 \eta\,(\ell+1)(\ell +2)^2}{R^2}\, \frac{\Phi_{\ell m}^{\rm out}}{R^{\ell+1}}
\nonumber\\
&&{} -\Bigl(2\,\frac{\kappa}{R^4}\, (\ell +2)^2(\ell -1)^2 + 2\,
\frac{K}{R^2}\, \ell  (\ell +1)+8\frac{K}{R^2}\Bigr)\,w_{\ell m}\nonumber\\
&&{} +\frac{1}{R^3}\Bigl(\,(K+\mu)\,\ell^2(\ell +1)^2-2\,\mu\, \ell (\ell +1)\nonumber \\
&&{} +4\,K\,\ell (\ell +1)\,\Bigr)\,\Psi_{\ell m}=0.
\label{nabla-expanded}
\end{eqnarray}

We use Darcy's law to set the normal velocity difference between the
fluid and the membrane, Eq.~(\ref{flux}). This generates an ordinary
first-order differential equation.  We look for solutions having a
time dependence of the form $\sim e^{-i \omega_{\ell m} t}$. Using
this assumption in Eq.~(\ref{flux}), we find our sixth and the last
relation between the dynamical variables,
\begin{eqnarray}
&&\Bigl[\,i\omega_{\ell m}\alpha-\left(\frac{\kappa}{R^4}(\ell +2)^2(\ell -1)^2+
4\frac{K}{R^2}\right)\Bigr]w_{\ell m}\nonumber\\
&&{}+\alpha\,\frac{\ell R}{2\eta (2\ell +3)} R^{\ell}P^{\rm in}_{\ell m}
+\alpha\,\frac{\ell}{R}\,R^{\ell}\Phi^{\rm in}_{\ell m}\nonumber\\
&&{}+2\,\frac{K}{R^3}\,\ell (\ell +1)\Psi_{\ell m}=0.
\label{Darcy-expanded}
\end{eqnarray}

We now have a set of six algebraic and differential equations
governing the six dynamical variables ($\Psi,w,\Phi^{\rm in},P^{\rm
in},\Phi^{\rm out},P^{\rm out}$). These six variables are governed
by Eqs.~(\ref{one}), (\ref{two}), (\ref{three}),
(\ref{normal-expanded}), (\ref{nabla-expanded}) and
(\ref{Darcy-expanded}). In this a damped
system, we expect to find solutions having a negative imaginary part
of $\omega_{\ell m}$, which we will interpret as the decay rate of a
mode describing the coupled dynamics of the spherical membrane and
fluid system in the case of vanishing in-plane membrane shear and
fluid vorticity.

Since we want to focus on the dynamics of the membrane, we use
Eqs.~(\ref{one}), (\ref{two}), (\ref{three}),
(\ref{normal-expanded}) to eliminate the four variables $p, \Phi$
associated with the fluid dynamics inside and outside the membrane.
As a result we find a system of two algebraic equations for
$w_{\ell m}$ and $\Psi_{\ell m}$ describing the compression and
bending of the membrane respectively:
\begin{equation}
\label{eq:w-psi}
A\begin{pmatrix} w_{\ell m}\\ \Psi_{\ell m}
\end{pmatrix}=0.
\end{equation}
The $\omega_{\ell m}$-dependent components of the $A$ matrix are
listed in Appendix \ref{A-mat-app}. The condition for the existence
of nontrivial solutions to Eqs.~(\ref{eq:w-psi}) determines the
decay rates of the deformation modes of the membrane coupled to the
surrounding fluid.

To discuss these decay rates of the membrane deformation modes, it
is convenient to introduce the following dimensionless variables
\begin{equation}
\label{reduced_parameters}
\displaystyle \bar{\omega}_\ell\equiv \frac{\omega_\ell\,\eta\,R^3}{\kappa},\
\displaystyle\bar{\mu}\equiv \frac{\mu \,R^2}{\kappa},\
\displaystyle\bar{\alpha}\equiv \frac{\alpha\,R}{\eta},\ \displaystyle \bar{K}\equiv\frac{K
R^2}{\kappa}
\end{equation}

\subsection{The modes of the shell}
\label{sub:modes}

\begin{figure}[htpb]
\centering
\includegraphics[width=7cm]{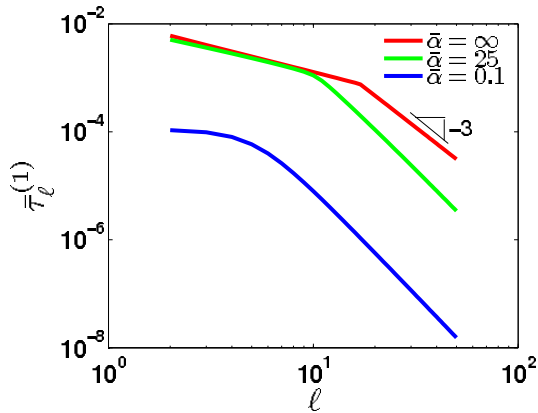}
\caption{(color online) Dimensionless relaxation times of the Lennon Brochard mode
$\bar{\tau}^{(1)}_\ell$ for an elastic membrane  vs the mode number
$\ell$ for various permeation parameters $\bar{\alpha}$. The
dimensionless elastic parameters of the membrane are $\bar{\mu}=75$
and $\bar{K}=225$, which corresponds to the Poisson ratio
$\sigma=0.5$ and the radius to thickness ratio $R/h=5$.
$\bar{\tau}^{(1)}_{\ell}=0$ for $\bar{\alpha}=0$ for all $\ell$.}
\label{f:tau1}
\end{figure}

In Fig.~\ref{f:tau1} we plot the decay time
$\tau^{(1)}_\ell=1/i \omega^{(1)}_\ell$ of the first mode as a
function of the order of the spherical harmonic $\ell\geq 1$. To
demonstrate the effect of permeation on these decay rates we show
three curves corresponding to a highly permeable membrane
($\bar{\alpha}=0.1$), one of intermediate permeability
($\bar{\alpha} = 25$) and a completely impermeable one
($\bar{\alpha}=\infty$).
All three cases correspond to a perfectly elastic membrane so that
$\bar{\mu}$ and $\bar{K}$ are real and frequency independent. For
$\ell \gg 1$ the characteristic wavelength of the deformation is
much smaller than the radius of curvature of the sphere so we should
expect to recover the flat membrane result of Lennon and
Brochard~\cite{Brochard:75}. Indeed, for large $\ell$, we find that,
$\bar{\tau}_{\ell}^{(1)}\sim \ell^{-3}$ so that the decay rate
$\left(\bar{\tau}_{\ell}^{(1)}\right)^{-1}
\sim\ell^3\sim\left(\frac{R}{\lambda}\right)^3\sim q^3$ where
$\lambda$ is the wavelength of the mode. As expected based on our
analysis of the plane permeable membrane, the transition to the
Lennon and Brochard behavior is more apparent for the impermeable
case ($\bar{\alpha}=\infty$) and becomes less distinct as
$\bar{\alpha}$ decreases. It is not possible to consider the case
$\bar{\alpha}=0$ when discussing the relaxational dynamics of the LB
mode as the decay rate of this mode becomes infinite as
$\bar{\alpha}\to 0$ for all $\ell$.

\begin{figure}[htpb]
\centering
\includegraphics[width=7cm]{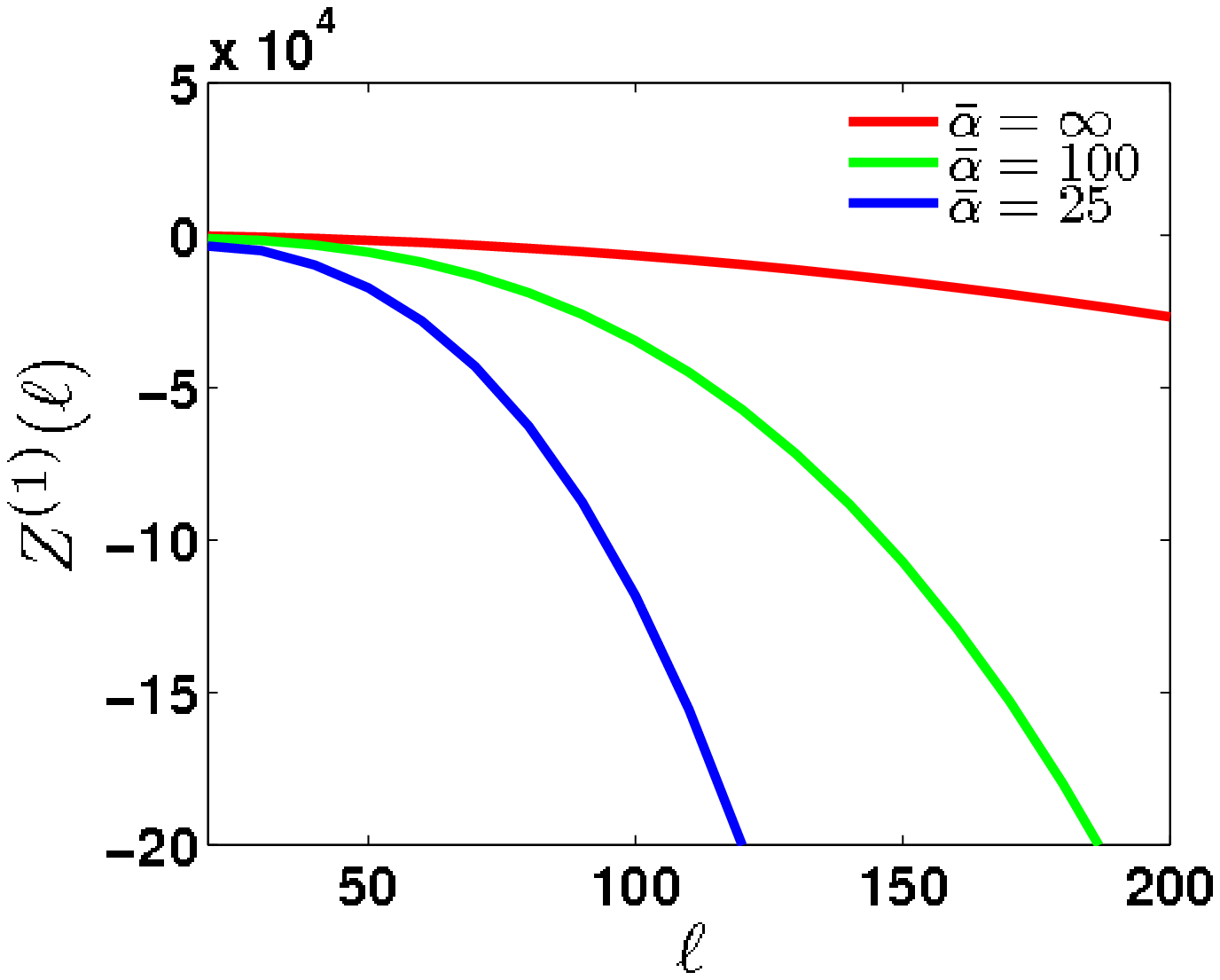}
\caption{(color online) Ratio of  bending to stretching $Z^{(1)}(\ell)$ in the
Lennon-Brochard mode vs. mode number $\ell$ of a perfectly elastic
membrane. The membrane elasticity parameters are identical to those used
in Fig.~\ref{f:tau1}. The values of the dimensionless permeation
coefficient are listed in the figure legend.} \label{f:Zmode1}
\end{figure}

We parameterize the relative degree of bending versus compression
associated with these modes by introducing their ratio
\begin{equation}
\label{mode}
w_{\ell m}/\Psi_{\ell m}=Z^{(p)}(\ell),
\end{equation}
where $p=1,2$ indexes the two linearly independent modes of the
system.  We plot in Fig.~\ref{f:Zmode1} this ratio $Z^{(1)}(\ell)$
for the first mode whose corresponding decay times are shown
in Fig.~\ref{f:tau1} . Examining this plot it is clear that this mode
is, indeed,  the analogue of the Lennon--Brochard bending mode since
$|Z^{(1)}(\ell)|\gg 1$. This identification of the first mode with
bending becomes more exact at higher $\ell$ where the curvature of
the membrane becomes less significant for the dynamics. The overall
negative sign for this result is due to the fact that where the
membrane bends outward the material expands and where it curves
inward the material compresses. Due to the overwhelming bending
character of this mode we will refer to it as the Lennon--Brochard
(LB) mode.

\begin{figure}[htpb]
\centering
\includegraphics[width=7cm]{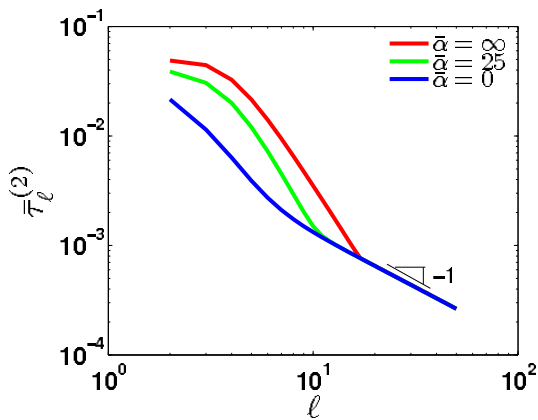}
\caption{(color online) Dimensionless relaxation time $\bar{\tau}^{(2)}_\ell$ in
the longitudinal sound mode vs. the mode number $\ell$ for purely
elastic membranes having various dimensionless permeation parameters
$\bar{\alpha}$. The dimensionless elastic parameters of the membrane
are $\bar{\mu}=75$ and $\bar{K}=225$ as in Fig.~\ref{f:tau1}.}
\label{f:tau2}
\end{figure}

We now turn to the second mode and plot in Fig.~\ref{f:tau2} its
dimensionless decay time $\bar{\tau}_{\ell}^{(2)}=1/i\omega^{(2)}_\ell$ as a
function of $\ell$. We use the same elastic parameters $\bar{\mu}$
and $\bar{K}$, as in the plot of the decay times
of the LB mode and show curves corresponding to an impermeable case
($\bar{\alpha}=\infty$), one of intermediate permeability
($\bar{\alpha}=25$), and a case where the fluid passes through the
membrane without any resistance ($\bar{\alpha}=0$).
When $\ell\gg 1$ all these cases converge onto a single curve so
that $\bar{\tau}_{\ell}^{(2)}\sim \ell^{-1}$ independent  of
$\bar{\alpha}$.

\begin{figure}[htpb]
\centering
\includegraphics[width=7cm]{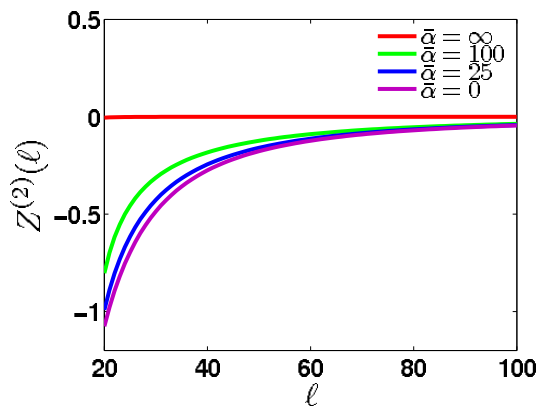}
\caption{(color online) Ratio of  bending to stretching $Z^{(2)}(\ell)$ in the
longitudinal sound mode vs. mode number $\ell$ of a perfectly
elastic membrane. The membrane elasticity parameters are identical
to used in Fig.~\ref{f:tau2}. The values of the dimensionless
permeation coefficient are listed in the figure legend.}
\label{f:Zmode2}
\end{figure}

As $\bar{\tau}_{\ell}^{(2)}\sim \ell^{-1}$ for large $\ell$, this
mode cannot be controlled by membrane bending at high wavenumber.
Turning to $Z^{(2)}(\ell)$ -- see Eq.~(\ref{mode}) --,
which we plot in Fig.~\ref{f:Zmode2}, we see that this mode is
dominated by the in-plane compression of the membrane. In fact, this
mode is the remnant of longitudinal sound (LS) in the elastic membrane
that is over-damped due to the coupling to the surrounding solvent.
It is interesting to note that only at large $\ell$ the (LB) mode
becomes strongly bending dominated while the longitudinal sound mode
(LS) only loses its bending character at large $\ell$.
It is well known that bending and longitudinal sound are orthogonal
normal modes of the linearized membrane/solvent system
in the case of {\em flat} membrane~\cite{Levine:02}. Examining the
results for the sphere we find that geometry couples these
deformations, but, at high wave number (or large $\ell$) where the
wavelength of the deformation is much less than the radius of
curvature $R$, the two orthogonal modes of the sphere approach the
simple mode structure of the flat membrane.

Before closing our discussion of the mode structure of the spherical
elastic membrane, we must consider the $\ell=0$ mode of the
membrane, which corresponds to purely radial motion of its surface.
This mode must be handled independently of the others because of the
incompressibility of the solvent. Because of that incompressibility,
there can be no corresponding radial motion of the solvent so that
the membrane relaxes against a background of quiescent fluid. 
In this mode the radial deformation obeys
\begin{equation}
\label{w-mode0}
\frac{\partial w}{\partial t}=-\frac{k_{\rm eff}}{\alpha}\, w,
\end{equation}
with no angular dependence, and with an effective spring constant of
\begin{equation}
\label{keff-mode0}
k_{\rm eff}=4 \bigg[\frac{\kappa + R^2 K }{R^4}\bigg ].
\end{equation}
The dimensionless relaxation rate of the $\ell = 0$ LB mode is then
given by Eqs.~(\ref{w-mode0}) and (\ref{keff-mode0}). We find
\begin{equation}
\bar{\tau}_0^{(2)}=\frac{\bar{\alpha}}{4}\,\frac{1}{1+\bar{K}}.
\end{equation}
It is clear that the decay rate of this mode vanishes in the limit
of an impermeable ($\bar{\alpha}=\infty$) sphere as volume changing
deformations are not allowed. As the sphere becomes more permeable
the decay rate increases becoming infinite in the limit of a ``ghost
membrane'' ($\bar{\alpha}=0$) since we have neglected the membrane's
inertia.

To compare these results with experiments it is useful to recognize
that our two-dimensional solid membrane is an idealization of a thin
sheet of thickness $h$ whose material properties can be defined in
terms of {\em three-dimensional} Lam\'e coefficients $\mu_3$,
$\lambda_3$. Using standard results \cite{Landau:59, Komura:92} we may
then  reexpress our dimensionless elastic parameters in terms of the
geometry of the system and the Poisson ratio of the membrane
material $\sigma$. We find
\begin{eqnarray}
\label{dimensionless-mu}
\bar{\mu}&=&6\left(\frac{R}{h}\right)^2(1-\sigma)\\
\bar{K}&=&6\left(\frac{R}{h}\right)^2(1+\sigma),
\label{dimensionless-K}
\end{eqnarray}
while the bending modulus is given by
\begin{equation}
\kappa=\frac{\mu_3 h^3}{3}\,\frac{\lambda_3+\mu_3}{\lambda_3+2\mu_3}
\end{equation}
We return to this point with regard to the microrheology of
viscoelastic spherical membranes in section~\ref{sec:response-function}.

\section{The nanoindentation response function}
\label{sec:response-function}

Using the above analysis of the mechanics of porous, viscoelastic
spherical shells, we now determine the
microrheological signature of viscoelastic, porous shells. As
discussed above, these results are relevant to the study of the
mechanics of viral capsids, colloidosomes, and porous lipid bilayers such as
those that contain  membrane-bound pore-forming proteins.

In the active AFM-based mechanical measurement, the shell is
deformed by antipodally applied forces
directed radially toward the center of the shell. To measure the mechanics
of the shell the distance between the point of force application and its
antipode is measured. This distance, due to the application of a
sinusoidally varying force $F_0 e^{-i\omega t}$, will deviate from the
shell diameter with an amplitude
\begin{equation}
\label{response-def}
D(\omega)=2R-H(\omega) F_0,
\end{equation}
which defines the response function $H(\omega)$. In the passive
microrheological experiment, on the other hand, the thermally excited
fluctuations of $2R-D$ give the same information via the
fluctuation-dissipation theorem~\cite{Mason:95,Levine:00}.

To calculate $H(\omega)$ we apply equal and opposite forces to the ``north'' and ``south''
poles of the sphere,
i.e. at polar angles $\theta=0,\pi$. The force balance
boundary condition Eq.~(\ref{F-normal}) then becomes
\begin{equation}
\hat{\boldsymbol{r}}\cdot \boldsymbol{F}_{\rm fluid}+\boldsymbol{\pi}_1{(\omega)}+
\boldsymbol{\pi}_2{(\omega)}=\hat{\boldsymbol{r}}\cdot \boldsymbol{F},
\end{equation}
where $\boldsymbol{\pi}_{1,2}$ are the external normal stresses on the surface
\begin{eqnarray}
&\boldsymbol{\pi}_1(\omega)&=-\delta(\theta) e^{-i\omega t}\, \hat{\boldsymbol{r}}\, F_0/a^2\\
&\boldsymbol{\pi}_2(\omega)&=\delta(\theta-\pi) e^{-i\omega t}\, \hat{\boldsymbol{r}}\, F_0/a^2.
\end{eqnarray}

As shown above we will consider point forces so that
the area of the sphere contacted by the AFM tip $a\to0$ at constant
$F_0/a^2=\pi_0$. Using the completeness of spherical harmonics
on the unit sphere we express these externally applied
stresses in terms of spherical harmonic modes:
\begin{equation}
\label{external_force}
\boldsymbol{\pi}(t,\theta,\phi)=-\hat{\boldsymbol{r}}\,\pi_0\sum_{\ell,m}
\left[ Y^*_{\ell 0}(0,0)+Y^*_{\ell 0}(\pi,0) \right]Y_{\ell m}(\theta,\phi)e^{-i\omega t}.
\end{equation}

Thus the AFM-induced pinching of the shell couples to
all the spherical harmonic modes of the system discussed above.
The effect of the finite-size of the AFM tip can be accounted for by 
putting in a large-$\ell$ cutoff in the sum shown in 
Eq.~(\ref{external_force}). For reasonably small tip, this 
effect must be negligible. Using these previous results we 
compute the radial deformation
amplitude associated with each $\ell-$mode at the north pole
of the sphere. In terms of the dimensionless frequency, $\bar{\omega}=\omega\eta R^3/\kappa$,
this radial displacement of the sphere at the north pole is given by
\begin{equation}
\label{north_pole}
w_{\ell}(\bar{\omega})=2\frac{R^4}{\kappa}\pi_0 [1+(-1)^{\ell}]Y^*_{\ell 0}(0,0)
\zeta_{\ell}(\bar{\omega}),
\end{equation}
where $\zeta_{\ell}(\bar{\omega})$ is the
frequency-dependent complex radial
compliance of the shell due to a radially applied normal stress
distributed over the sphere as $Y_{\ell 0}(\theta)$.
This compliance results from both the Lennon-Brochard (LB)and
longitudinal sound (LS) modes of the shell discussed above. The complex,
frequency-dependent response function can be  expressed in
terms of five dimensionless functions of the angular
momentum of the mode $\ell$:
\begin{equation}
\label{response_function}
\zeta_\ell(\bar{\omega}) = \frac{\Lambda_1(\ell)+i\,\bar{\omega}\,\Lambda_2(\ell)}
{\Lambda_3(\ell)+i\,\bar{\omega}\,
\Lambda_4(\ell)+\bar{\omega}^2\,\Lambda_5(\ell)},
\end{equation}
where due to their length, these
functions $\Lambda_1(\ell), ..., \Lambda_5(\ell)$ are reported
in Appendix~\ref{response-app}.
As shown there, it is through these functions that the
$\ell-$dependent response function acquires its
dependence on the dimensionless permeation parameter
$\bar{\alpha}$ and on the elastic constants of the membrane
written in terms of the ratio $R/h$ and the Poisson ratio $\sigma$
as shown in Eqs.~(\ref{dimensionless-mu}) and (\ref{dimensionless-K}).

Summing over the $\ell-$~modes
using Eqs.~(\ref{external_force}) and (\ref{north_pole}) we find the
radial displacement of the shell at $(\theta,\phi)$ to be
\begin{equation}
w(\bar{\omega},\theta,\phi)=\frac{\pi_0 R^4}{\kappa}\sum_{\ell=0}^{\infty}
Y_{\ell 0}(\theta,\phi)[1+(-1)^{\ell}]\zeta_{\ell}(\bar{\omega}) Y^*_{\ell 0}(0,0).
\label{radial-displacement}
\end{equation}
Thus the amplitude of oscillations of the diameter is given by
\begin{equation}
\label{D-w}
D(\bar{\omega})=2R+w(\bar{\omega},0,0)+w(\bar{\omega},\pi,0).
\end{equation}
Using the definition in Eq.~(\ref{response-def}) and 
Eqs.~(\ref{response_function}), (\ref{radial-displacement}), 
and (\ref{D-w} ), we solve
for the response function in terms of the point
force response of the shell. We find that
the diameter response function is given by
\begin{equation}
H(\bar\omega)=\frac{2R^4}{\kappa}\sum_{\ell=0}^{\infty}[1+(-1)^{\ell}]
\zeta_{\ell}(\bar{\omega}) |Y_{\ell 0}(0,0)|^2.
\end{equation}

\subsection{An elastic shell}
\label{sub:results-elastic}
\begin{figure}[htpb]
\centering
\includegraphics[width=7cm]{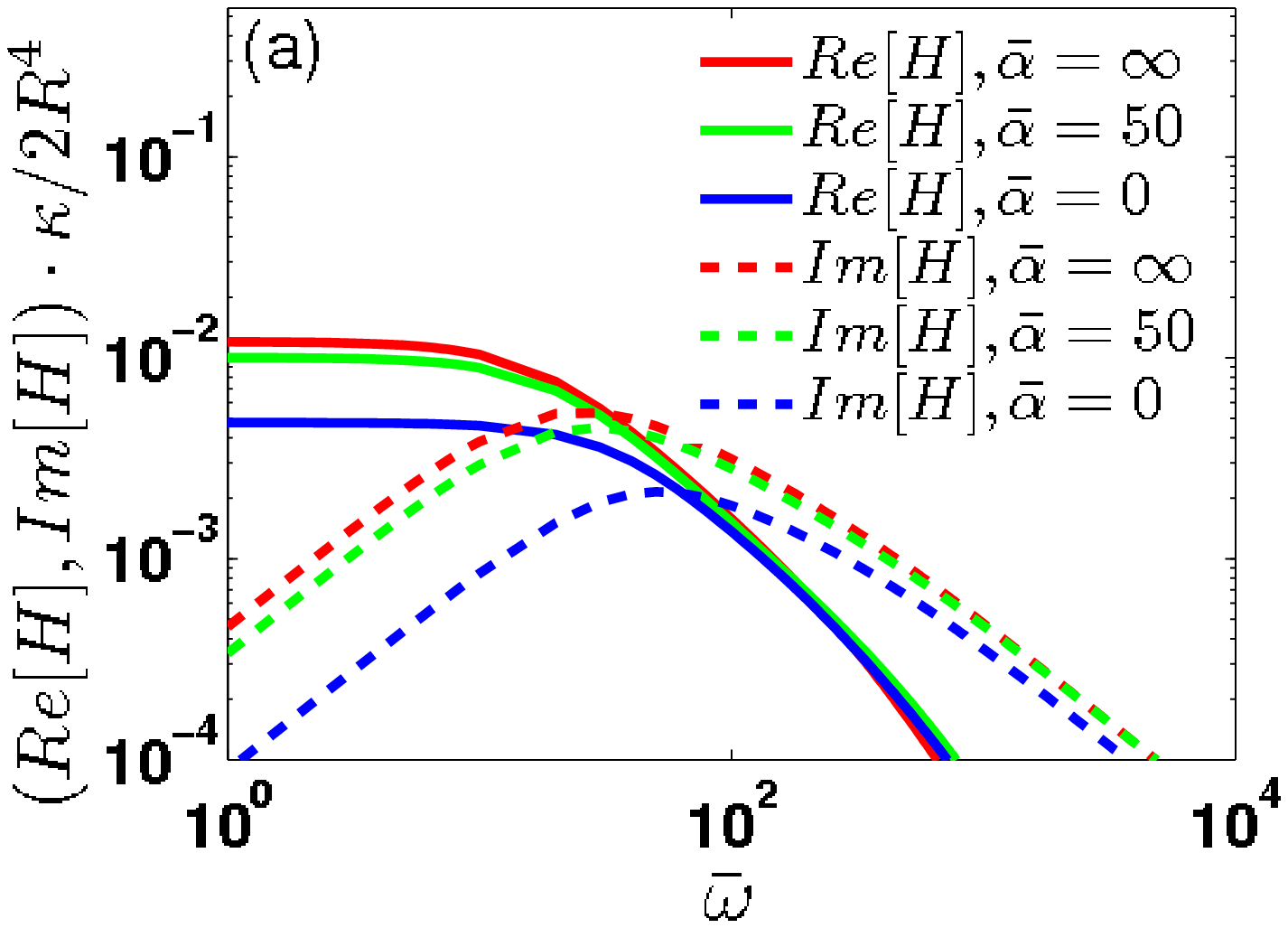}
\includegraphics[width=7cm]{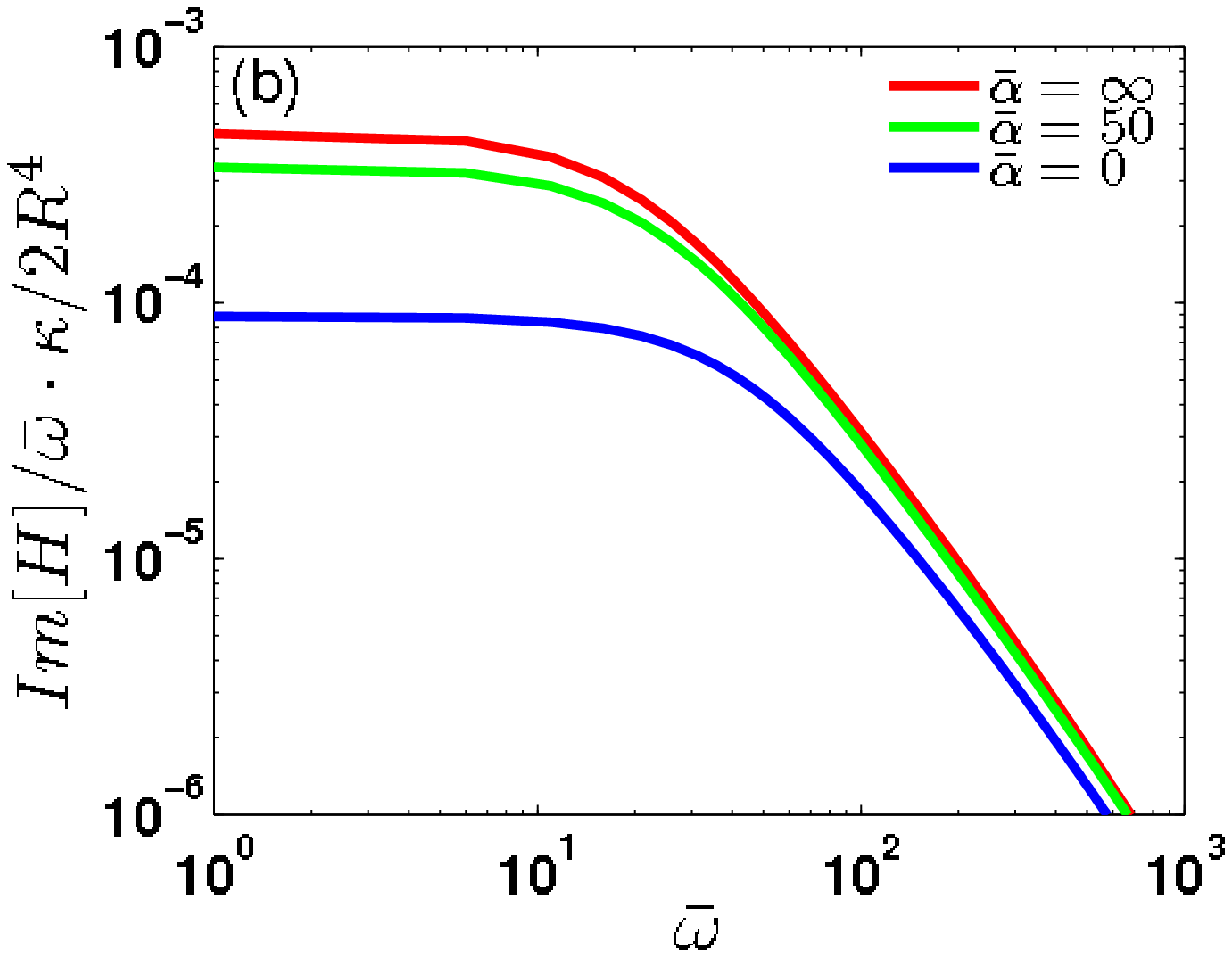}
\caption{(color online) The response function of a porous, elastic shell.
In panel (a)
we plot the real (${\rm Re}[H(\bar\omega)]$, solid lines)
and imaginary
(${\rm Im}[H(\bar\omega)]$, dashed lines)
parts of the response function vs dimensionless frequency $\bar{\omega}$ for various values
of the permeation parameter  $\bar{\alpha}$.
The remaining material parameters are: $R/h=5$,
Poisson ratio $\sigma=0.5$.
In panel (b) we show the thermal power spectrum
(${\rm Im}[H(\bar\omega)]/\bar\omega$) using the same material parameters.}
\label{fig:elastic}
\end{figure}
We plot in Fig.~\ref{fig:elastic}(a) the real (in phase) and imaginary
(out of phase) parts of the
diameter response function $H(\bar{\omega})$ for purely elastic shells of
varying porosity. In anticipation of future fluctuation-based microrheology
experiments we plot in Fig.5(b) the  predicted power spectra for the thermal
fluctuations of this variable, computed
using the fluctuation-dissipation theorem \cite{Landau:51}.  Examining the figure we note
that the response function can be characterized as having a low-frequency elastic plateau
followed by a cross-over to a viscous-dominated decay at high frequencies. As the dimensionless
permeation coefficient is varied from that of an impermeable shell ($\bar{\alpha} = \infty$) to one that
allows fluid permeation with no resistance ($\bar{\alpha} = 0$), the dimensionless cross-over
frequency moves from $\approx 10$ to $\approx 10^2$. We return to predicted values of this
cross-over frequency for various physical systems in section~\ref{sec:discussion}.  Within the
low-frequency plateau where much, if not all, the dynamical data is likely to be taken, the main
effect of permeation is to shift the value of the response function by approximately a factor of two.
Thus, in order to correctly extract the bending and compression moduli of the shell to better than
this factor of two, an accurate account of the permeation must be made. In the passive
microrheology experiment, the effect of permeation on the thermal power spectrum of diameter fluctuations
is more pronounced. Ignoring permeation in this sort of measurement leads to quantitative inaccuracies on
the order of $4$ as shown in Fig.~\ref{fig:elastic}(b). We now consider the case of a
viscous shell.

\subsection{A viscous shell}
\label{sub:results-viscous}
The case of a viscous shell can be physically realized with giant
unilamellar vesicles (GUVs) that contain pore-forming transmembrane proteins or protein complexes.
This system allows for the largest variation of permeation coefficients of any that we study.
By varying the number of such proteins in these protein/lipid complexes,
one may vary the porosity parameter from essentially
$\bar{\alpha} =\infty$ (no pore-forming protein) to finite values.  
Using the solution for the permeation coefficient found in Appendix~\ref{Appendix-A},
it should be possible to reach values of $\bar{\alpha}=10$ for the lipid vesicle systems.

To determine the dynamics of a
purely viscous shell we replace real, frequency-independent shear modulus of the shell by a
viscous response: $\mu\to -i\omega \eta_{\rm m}$, with $\eta_{\rm m}$ being the
two-dimensional surface viscosity of the the membrane or bilayer. 
To allow any radial deformations of the impermeable bilayer, there must be
some area change of the surface. One may imagine that the
surface of a tensed GUV is essentially inextensible, however, even such
\begin{figure}[htpb]
\centering
\includegraphics[width=6cm]{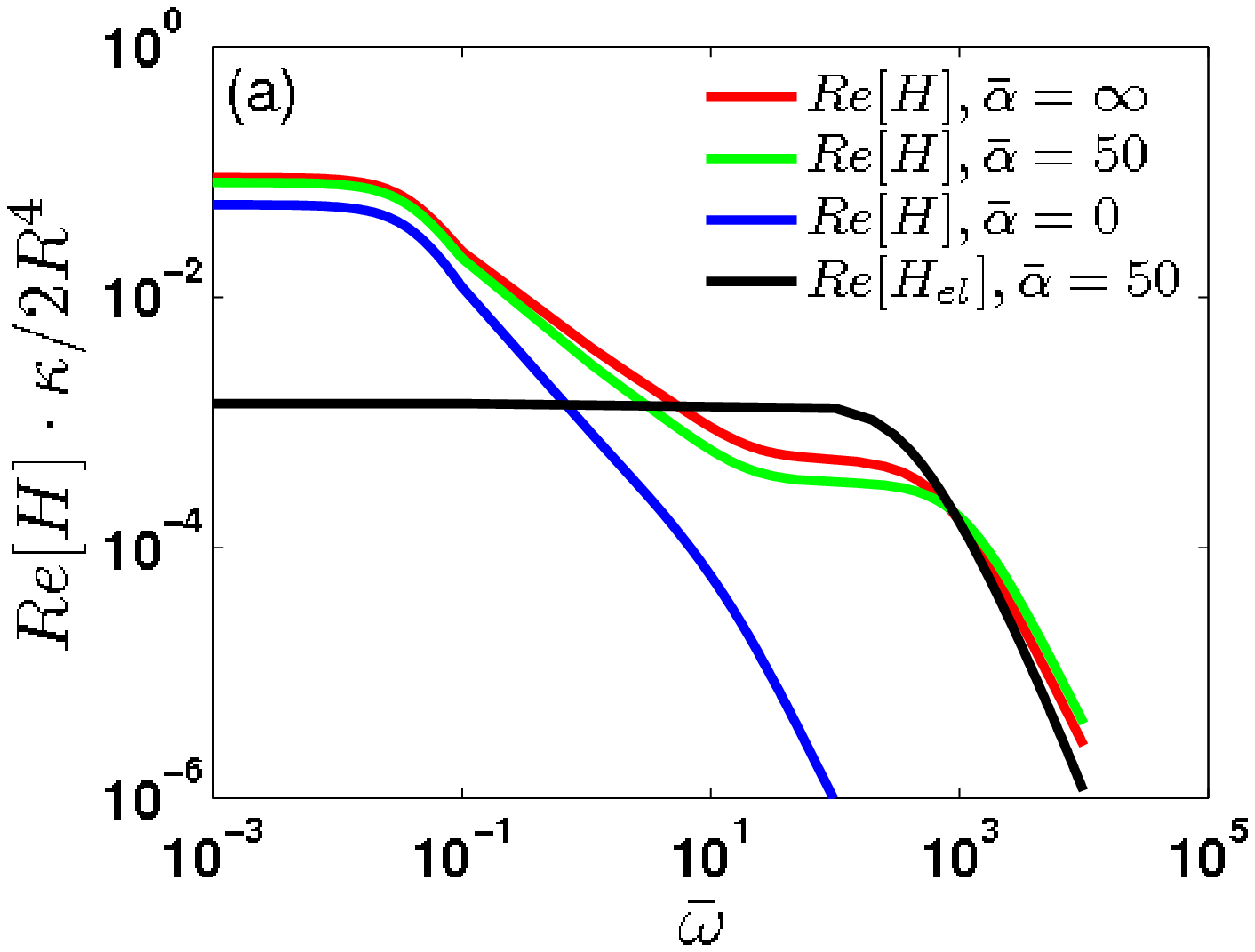}
\includegraphics[width=6cm]{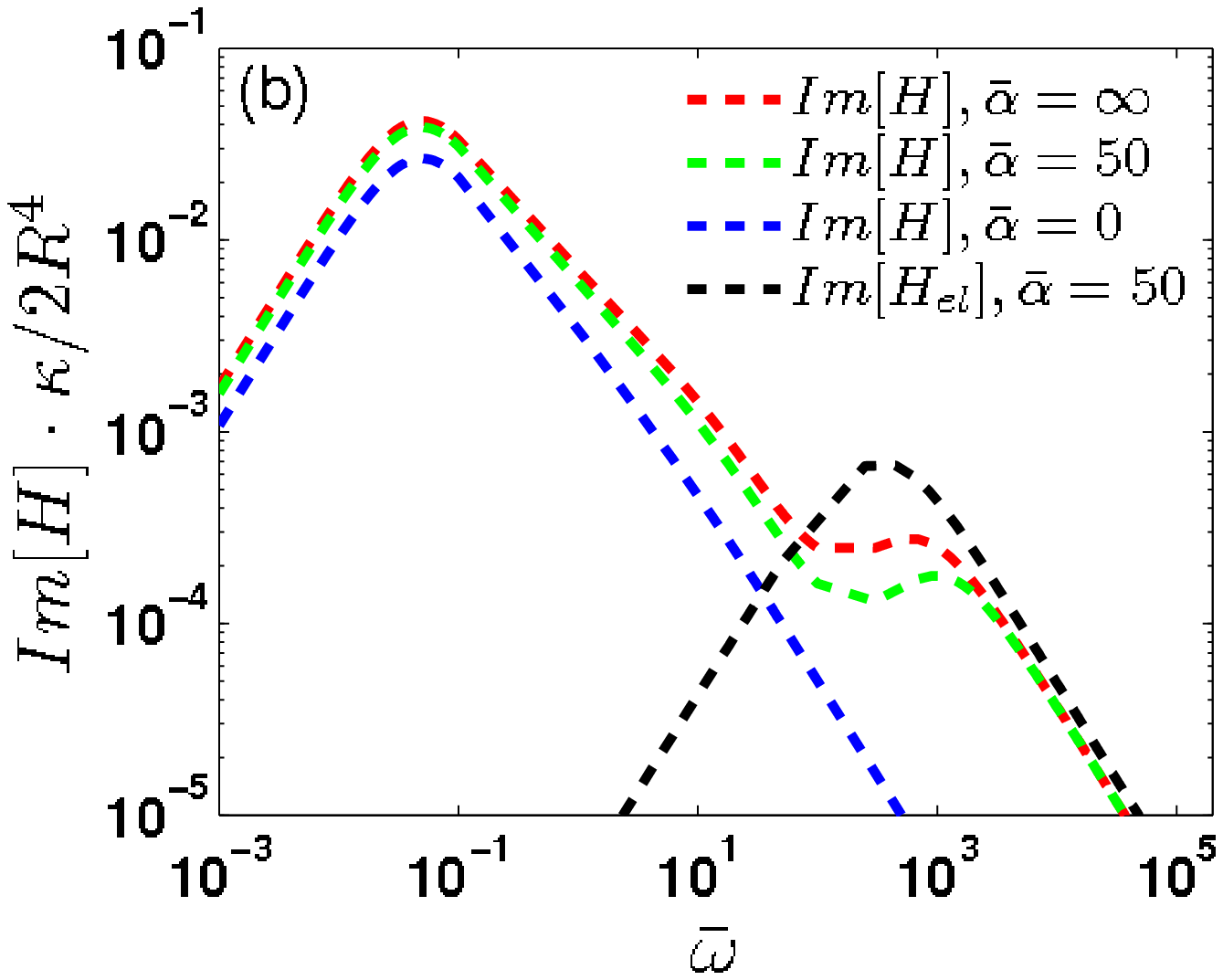}
\includegraphics[width=6cm]{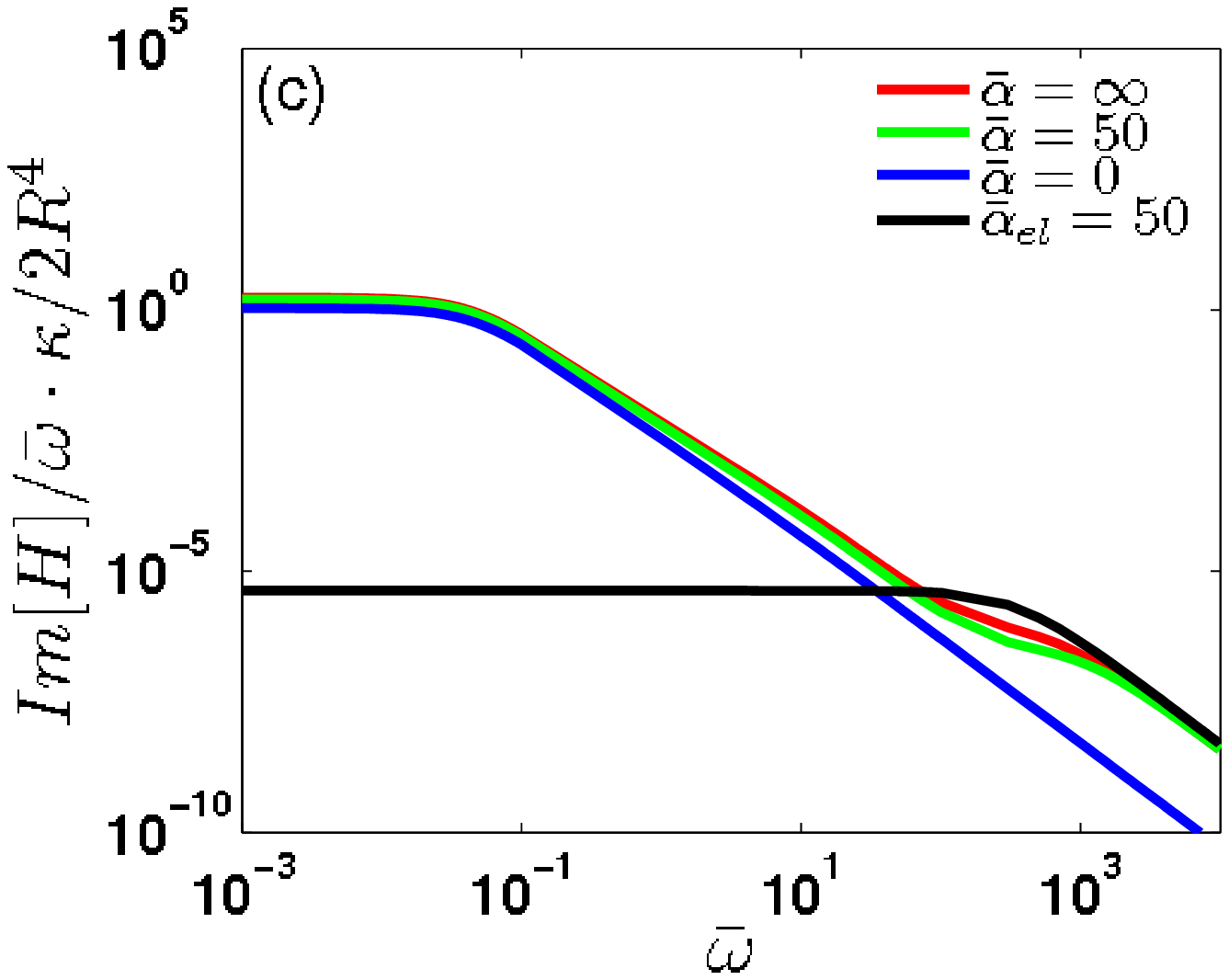}
\caption{(color online) Response function of a GUV containing pore-forming
protein complexes. In panels (a) and (b) we show the real (${\rm Re}[H(\bar\omega)]$)
and imaginary (${\rm Im}[H(\bar\omega)]$) parts of the response function
vs. dimensionless frequency $\bar{\omega}$ respectively using
various values of the permeation parameter $\bar{\alpha}$.
The other dimensionless parameters were:$R/h=20$, $\sigma=0.5$
$\eta_{\rm m}/R\eta=0.1\bar{\mu}_0$, $\bar{\mu}_0$ and $\bar{K}$ defined 
in Eqs.~(\ref{dimensionless-mu}) and (\ref{dimensionless-K}). For comparison, 
black solid line shows the real part of the response function for an 
elastic shell with $R/h=20$, $\sigma=0.5$, $\bar{\alpha}=50$.
The black dashed line corresponds to the imaginary part of the response function 
of an elastic shell with the same material parameters as in panel (a).
In (c) we plot the thermal power spectrum (${\rm Im}[H(\bar\omega)]/\bar\omega$)
as a function of the dimensionless frequency $\bar\omega$ using the
same material parameters as in panel (a). Black line -- elastic shell.}
\label{fig:GUV}
\end{figure}
tensed structures contain a reservoir of extra area hidden in the
(small-scale) thermal undulations of the surface~\cite{Milner:87, Kaizuka:04}.
Thus the two-dimensional bulk modulus of the viscous shell is dominated by a large,
but finite elastic component. We expect that
there may well be a dissipative
or viscous response of the shell to
area-changing deformations, but we
ignore these subdominant corrections.
In Fig.~\ref{fig:GUV}(a), we plot the real and imaginary
parts of the response function for a porous ($\bar{\alpha}=50$),
purely viscous shell with a finite area compression
modulus, $\bar{K}$.  In part (b) of the same figure,
we calculate the power spectrum of thermal
fluctuations of the diameter of such an object.

The low frequency response of the porous, viscous membrane exhibits an elastic plateau due to its elastic response to 
both bending and compression. The viscous shell, however, admits a new intermediate frequency
response regime associated with viscous dissipation \emph{within} the fluid membrane. In this intermediate 
frequency regime, the real part of the response function decays as $\bar{\omega}^{-0.66}$ over about 
two decades in frequency. Examining these results we note that there are dramatic qualitative differences between the 
response functions of viscous and elastic shells at least within this intermediate frequency range. 

In Fig.~\ref{fig:GUV}(b) we see the low-frequency peak in the imaginary part of the response function associated with 
these internal (to the membrane) shear modes. At still higher frequencies, both the real and imaginary
parts of the response function are dominated by dissipative stresses due to solvent flow and permeation 
through the pores. This can be most easily seen in the correspondence of the high-frequency peak of the
imaginary part of response function of the viscous membrane with only peak in the corresponding part of 
the response function for the purely elastic shell. In this high frequency limit
a viscous membrane and an elastic shell show essentially an identical mechanical response. This is reasonable 
since the high-frequency dynamics of both systems are dominated by solvent flow and permeation. 

Passive microrheological measurements at low frequencies can easily distinguish between a viscous and an elastic
shell as can be see by the difference between their power spectra shown in Fig.~\ref{fig:GUV}(c). The effect of 
permeation, as can be seen by comparing the curves, is principally in shifting the hight of the 
plateau region and slightly altering the transition frequency to the solvent-dominated, high-frequency regime.

\subsection{Viscoelastic shells}
\label{sub:results-viscoelastic}
To explore the role of viscoelasticity in the mechanics of a porous
spherical shell, we consider two archetypal
cases.  We first calculate the response function for
a viscoelastic shell where the shear modulus has a single relaxation time, $\tau_s$.
This Maxwell model~\cite{Bird:77} of the shear modulus can be written as
\begin{equation}
\label{maxwell}
\bar\mu(\bar\omega)=\bar{\mu}_0\,\frac{-i\bar\omega\bar{\tau}_s}{1-i\bar{\omega}\bar{\tau}_s},
\end{equation}
where $\bar{\tau}_s$ is the dimensionless shear stress relaxation time. Deformations
occurring at a frequency scale much
lower than the inverse stress relaxation time relax viscously. The high frequency dynamics,
on the other hand, see an elastic shear response of the
material with shear modulus $\bar{\mu}_0$.

\begin{figure}[htpb]
\centering
\includegraphics[width=6cm]{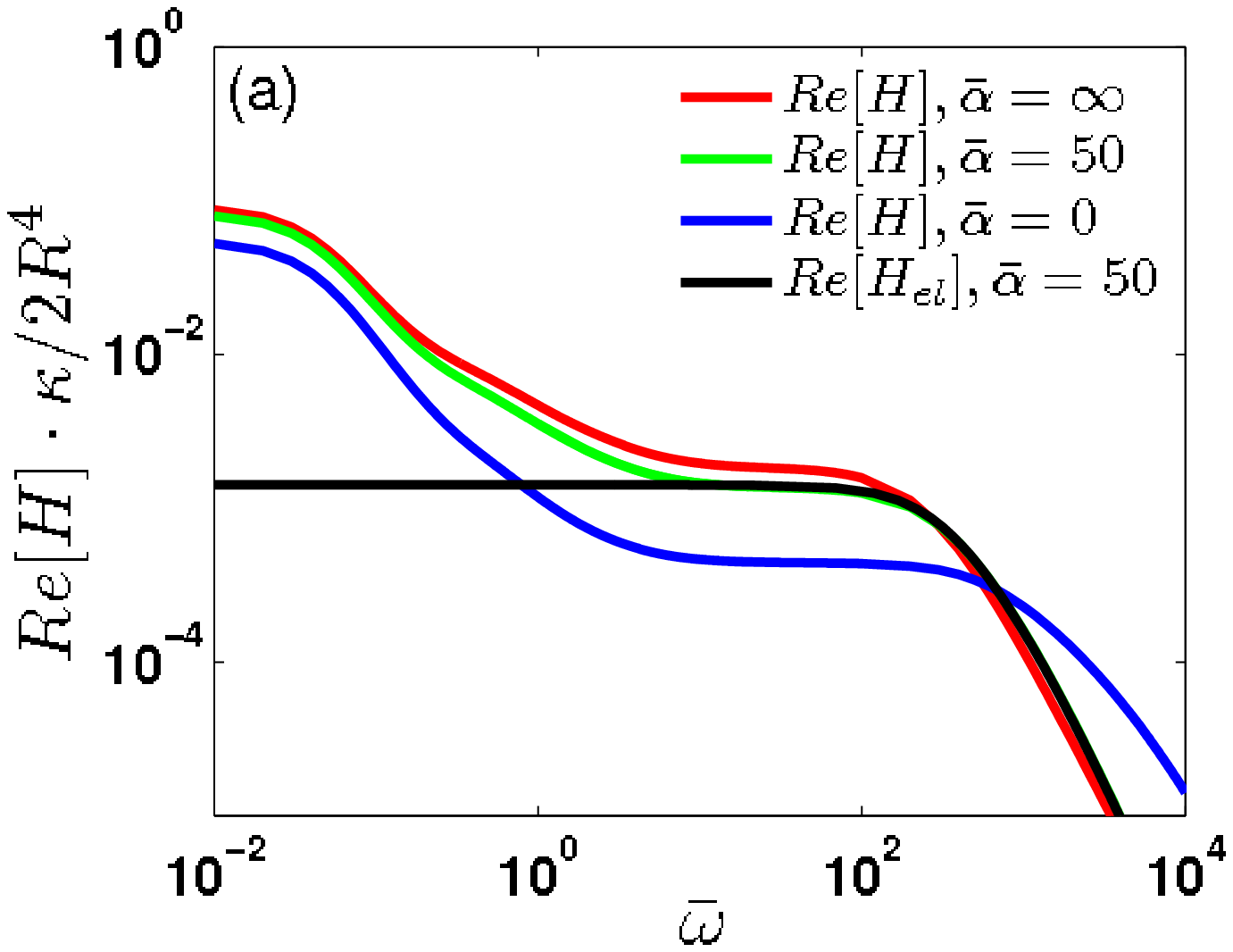}
\includegraphics[width=6cm]{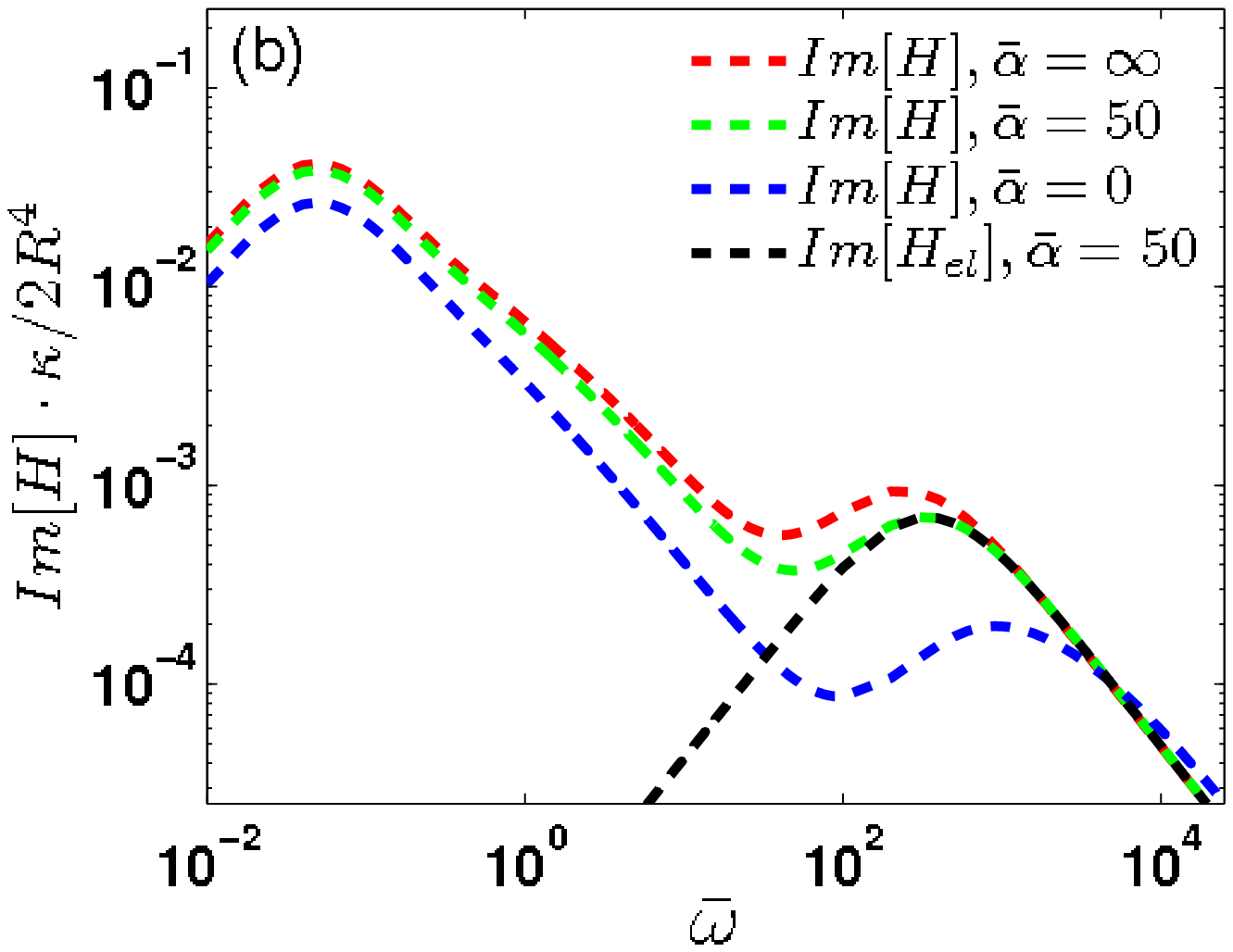}
\includegraphics[width=6cm]{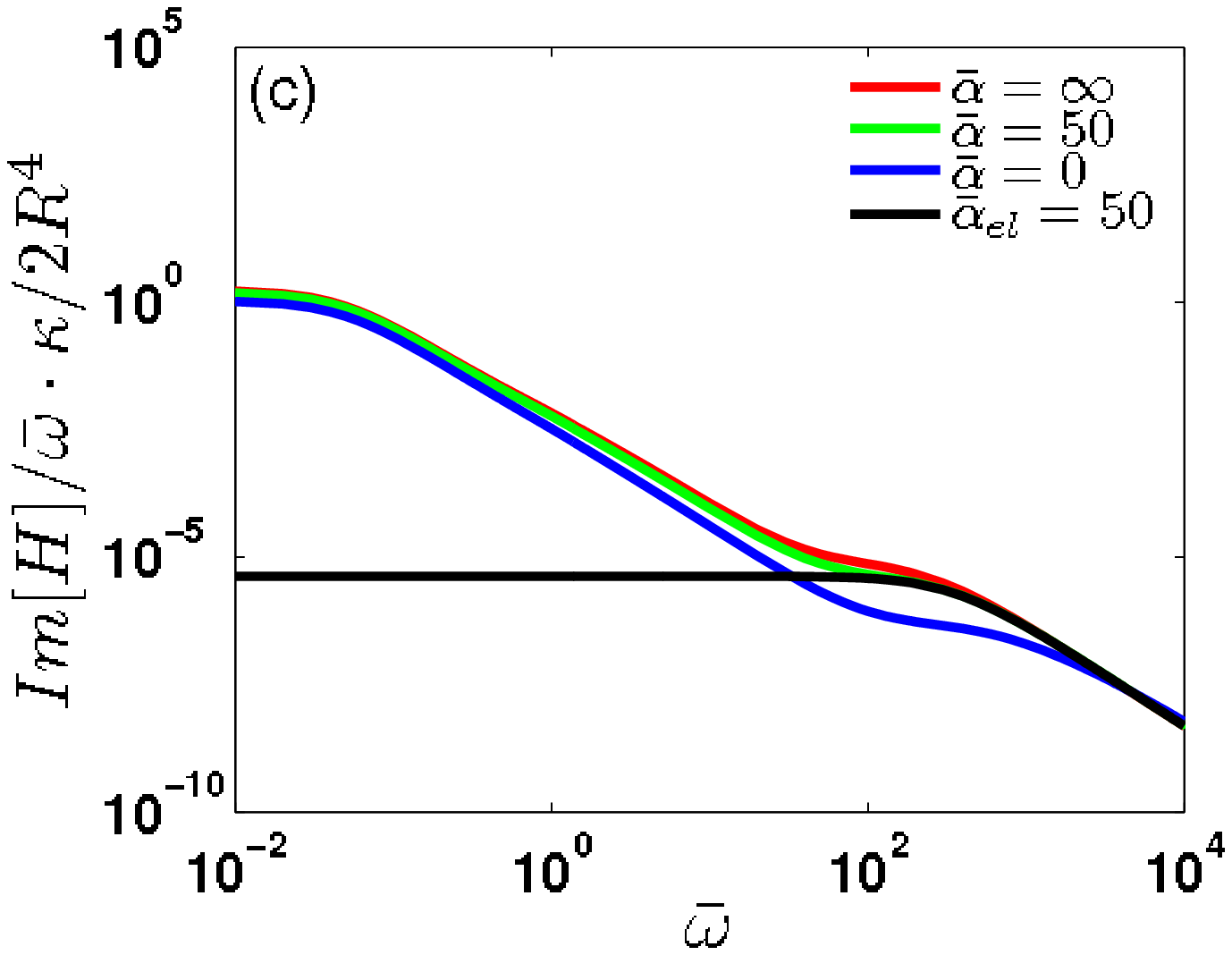}
\caption{(color online) The response function of
a shell with a viscoelastic shear modulus.
In panels (a) and (b) we show the real (${\rm Re}[H(\bar\omega)]$)
and imaginary parts of the response function
vs. dimensionless frequency $\bar{\omega}$ respectively using
various values of the permeation parameters $\bar{\alpha}$.
The other dimensionless parameters were: $R/h=20$, $\sigma=0.5$, $\bar\tau_s=0.1$,
$\bar{\mu}_0$ and $\bar{K}$ defined in Eqs.~(\ref{dimensionless-mu}) and (\ref{dimensionless-K}). 
For comparison, the black lines shows the real (solid) and imaginary (dashed) 
parts of the response function for an elastic shell with $R/h=20$, $\sigma=0.5$, $\bar{\alpha}=50$.
In (c) we plot the thermal power spectrum (${\rm Im}[H(\bar\omega)]/\bar\omega$)
as a function of the dimensionless frequency $\bar\omega$ using the
same material parameters as in panel (a). Black line -- purely elastic shell.}
\label{fig:viscoelastic}
\end{figure}

In Figs.~\ref{fig:viscoelastic}(a) and (b) we plot the real and
imaginary parts of the response function respectively
for a viscoelastic shell having a complex, frequency-dependent shear modulus
of the Maxwell form.
We take the shear stress relaxation time $\bar{\tau}_s$ to be within
the range of decay times of the
various $\ell-$modes of the system. Thus, the small
$\ell-$modes of the shell
have long decay times relative to the stress relaxation time and thus
experience viscous relaxation dynamics, where the higher $\ell-$modes
of the system decay fast enough to be affected by the elastic shear response of the system. 

It is that for a viscoelastic shell the intermediate frequency response regime associated with the 
viscous system remains, but the frequency dependence of the real part of the response function is 
weakened: $\sim \bar{\omega}^{-0.44}$. This appears reasonable as the viscoelastic case
effectively interpolates between the response of an elastic shell (high $\ell$-modes) and a viscous 
one (low $\ell$-modes).  The slope of the real part of the response function (see Fig.~\ref{fig:viscoelastic}(a))
in the intermediate response regime now is $\bar{\alpha}$-dependent. Thus, to use the response function to extract
properties of the viscoelastic response of the shell is difficult without an accurate model of solvent permeation 
through it. 

In Fig.~\ref{fig:viscoelastic}(c) we plot the thermal power spectrum for this model. As long as
the shear stress relaxation rate is small compared to the cross-over to the high-frequency terminal behavior
of the system, it is indeed possible to observe the viscoelastic response of the shell by examining the 
thermal power spectrum. This identification, however, relies on a separation of time scales 
so that $\bar{\tau}_s$ remains long compared
to the cross-over timescale to viscous-dominated behavior. As we discuss in section~\ref{sec:discussion},
this separation of time scales should be easy to achieve in many physical systems.

Finally, we study perhaps the most potentially interesting example of
a viscoelastic shell on the nanoscale.  In some viral capsids the individual
capsomeres can undergo a conformational change that changes their effective cross-sectional
area.  These internal degrees of freedom related to volume-changing allosteric
transitions of the proteins making up
the shell of the virus allow for new dissipation associated with
area-changing deformations. To model this system we include a
complex frequency-dependent two-dimensional bulk modulus of the form
\begin{equation}
\bar{K}(\bar{\omega})=\bar{K}_0\left[1+ g \frac{-i\bar{\omega}\bar{\tau}_B}
{1-i\bar{\omega}\bar{\tau}_B }\right],
\label{allosteric-K}
\end{equation}
where $\bar{K}_0$ sets the overall modulus scale, $\bar{\tau}_B$ is
the inverse rate of conformational change, and $g\bar{K}_0$ sets
the maximum of the dissipative part of the modulus at $\bar{\omega}\bar{\tau}_B\simeq 1$.
\begin{figure}[htpb]
\centering
\includegraphics[width=6cm]{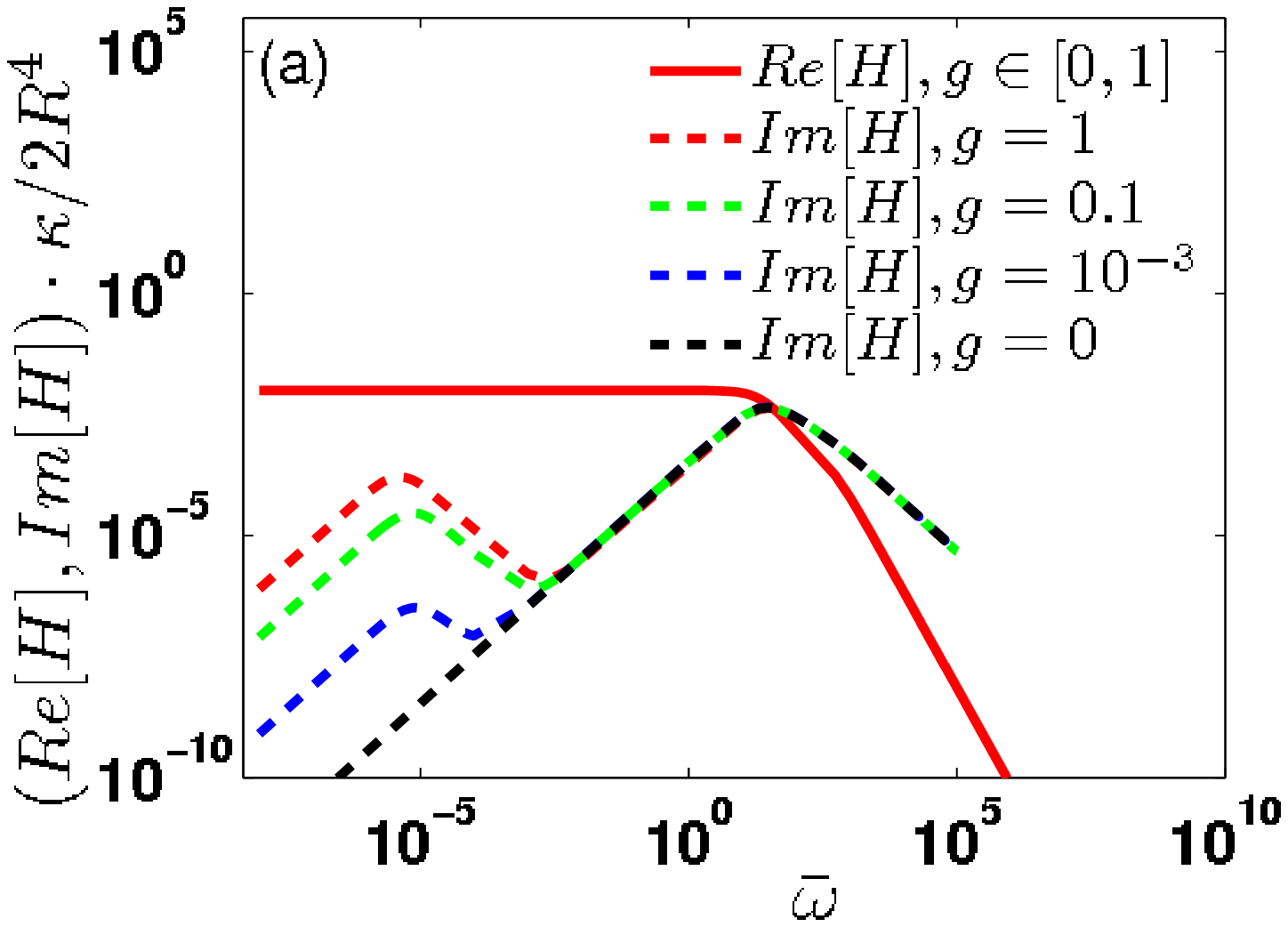}
\includegraphics[width=6cm]{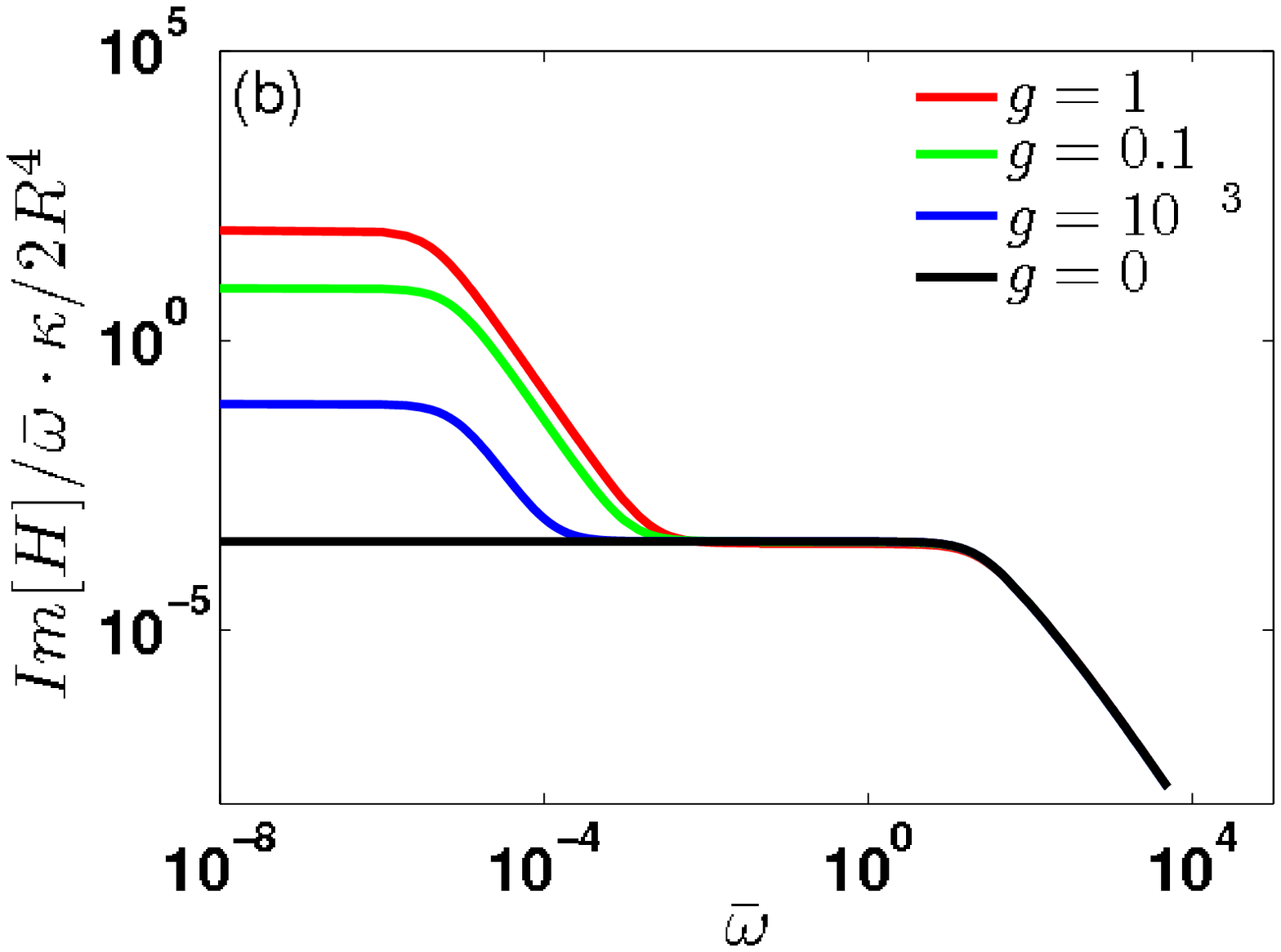}
\caption{(color online) A simple mechanical model of a
viral capsid in which the capsomers
undergo a cross-sectional area changing allosteric transition.
In panel (a) we plot the
real and imaginary parts of the response function for
different Maxwell solid models of the shell --
see Eq.~(\ref{allosteric-K}). In panel (b) we
plot the expected thermal power spectrum
(${\rm Im}[H(\bar\omega)]/\bar\omega$). In all cases the
material parameters of the shell are given by: $R/h=5$, $\tau_B=10^{-3}$s, $Y=140$MPa, $\sigma=0.5$, $\bar\alpha=50$.}
\label{fig:allosteric}
\end{figure}
The typical
time scale for allosteric transition in proteins is on the order of microseconds to milliseconds. The ratio of the
zero-frequency elastic part of the area modulus $\bar{K}_0$ to the dissipative contribution at the
frequency of these allosteric transitions is controlled by the constant $g$. Since we know of no quantitaive measurement of these parameters, we sweep the value of $g$ from zero, which corresponds to a purely elastic shell as shown in
Fig.~\ref{fig:elastic}, to $g=1$, where this dissipative contribution to the stress dominates the
zero-frequency elastic contribution to the area modulus. It is clear from Fig.~\ref{fig:allosteric} that
even a small dissipative response to compressive stresses at time scales associated with allosteric
transitions leads to a clear and measurable change in the response function of the shell.

\section{Summary and Discussion}
\label{sec:discussion}

We have developed a theoretical model for the dynamics of
a porous spherical shell immersed in a viscous solvent and used it to examine the
utility of nanoindentation-based rheological measurements.  In the limit of a flat,
porous membrane immersed in
a viscous solvent, we calculated the relaxational dynamics of the bending or undulatory
modes of the surface. These modes are
well-understood in the limit of a impermeable surface.
In that limiting case, the long-range
hydrodynamic flows introduce a nonlocal coupling of
the stress and displacement along the surface. It has long been known that the
hydrodynamic coupling shifts the dispersion relation of
the undulation modes of wavevector $q$ from their
expected ``local drag'' form of $\omega \sim - i q^4$ to correct answer of
$\omega \sim - i q^3$ incorporating the hydrodynamic cooperativity
of the system as first proposed by Lennon and Brochard~\cite{Brochard:75}. For the case of
permeable membrane, the dispersion relation of the undulatory modes smoothly
crosses over from the local drag form to the usual Lennon-Brochard
solution as the wave number is varied. The cross-over length is set by $\ell \sim \eta/\alpha$.
At lengths much larger than
$\ell$ the solvent permeation generates a subdominant correction to the usual
dynamics of the undulatory modes of a membrane. Below this length, however, permeation effectively
destroys the hydrodynamic coupling along the membrane leading to a new dynamics consistent with the
na\"{i}ve, local-drag approximation.

In order to consider the mechanics of spherical permeable membranes (of radius $R$), we
examined both compression and bending deformations of the material since curvature couples these linearly
independent modes of the flat membrane. The hydrodynamic interactions of the membrane are also
less simple in this geometry. Using a solution for the solvent dynamics both inside and outside the
spherical shell, we determined the over-damped normal modes of the system. Owing to the spherical symmetry
of the problem, these modes are proportional to the spherical harmonics and thus can be
indexed by the usual angular momentum variables, $(\ell,m)$. At small angular momentum, i.e. $\ell \sim 1$, the
two linearly independent modes of the membrane each combine an undulatory and compressional character. At
higher angular momentum corresponding to higher wavevectors on the surface $q \sim \ell/R$, the dynamics
approach those of the flat interface where these two modes acquire a dominantly undulatory or compressional
character respectively. Using this short length scale behavior, we label the
modes as either the Lennon-Brochard (i.e. undulatory) or compressional type.

We use these dynamical calculations to study the expected finite-frequency response of a variety of essentially
spherical, viscous, elastic, and viscoelastic porous shells.
Such systems include colloidosomes, nanoparticle
networks, vesicles containing membrane-bound pore-forming protein
complexes, and viruses.  The principal questions
that we address are how does porosity affect the mechanical response
of viscous, elastic, and viscoelastic spherical shells and specifically, for the case of viral capsids, is it possible
to mechanically detect allosteric transitions in the capsomeres making up the proteinous outer shell of a virus?

To address these questions, we need only determine the characteristic frequencies associated with
the nondimensionalized curves shown in Figs.~\ref{fig:elastic}, \ref{fig:GUV},
\ref{fig:viscoelastic}, and \ref{fig:allosteric}.  A few general comments are in order. First, the
characteristic frequency associated with the transition  between the
low-frequency plateau and high-frequency, fluid dominated decay of the response function is typically
large enough in viruses to render the high-frequency region unobservable using current methods. Expressing the
frequencies $f$ in dimensional form using the dimensionless angular frequency $\bar{\omega}$ and the
relation
\begin{equation}
\label{frequency-dimensions}
f=\bar{\omega} \left[\frac{Y}{24 \pi\,\eta (1-\sigma^2)}\right] (h/R)^3,
\end{equation}
we find that the characteristic cross-over frequency for a CCMV virus~\cite{Michel:06} 
to be on the order of $10^8$Hz. For a larger
and mechanically softer liposome, however, this transition frequency is $\sim 100$kHz. For colloidosomes
and other synthetic structures on the scale of one to ten microns we estimate this transition frequency to be
$\sim 10^2 - \, 10^4$Hz. The general trends are simple to understand
from Eq.~(\ref{frequency-dimensions}). Larger (large $R$) and thinner-walled (small $h$) spheres
have a lower transition frequency, while those spheres made of less elastically compliant materials
(i.e. higher Young's modulus $Y$) have higher transition frequencies.

We expect, based on the above estimates, that the transition to a permeation-dominated response is
not observable in viral capsids. One advantage of this point is that an observed increase in the
imaginary (out-of-phase) response function for viral capsids implies the existence of internal dissipative
modes of the structure that are likely to be related to allosteric transitions of capsomeres or in their
rearrangement relative to one another. We show in Figs.~\ref{fig:viscoelastic},\ref{fig:allosteric} two
possible rheological signatures of such internal dissipative modes.  In the former we suppose that there is a
large elastic response of the shell to area-changing deformations, but that the shear response of the material
is viscoelastic.  One might expect such a mechanical response if the individual capsomeres
irreversibly rearrange in response to applied stress. In the latter case we explore a system having a
Maxwell solid response to area compression, but
an elastic response to shear. Such a material response might be expected if the individual capsomeres undergo
allosteric transitions that change their effective cross-sectional area.
In Fig.~\ref{fig:viscoelastic}, we see that the Maxwell shear response leads to measurable corrections 
to the rheological spectrum of the object. 

The AFM-based mechanical probe is also highly
sensitive to a viscoelastic response of the area compression modulus as might be due to 
allosteric transitions in the capsomeres. The expected
separation of time scales between the lower frequency allosteric transitions ($10^{3}$Hz to $10^{6}$Hz) and
transition to permeation dominated dynamics ($10^{8}$Hz) demonstrates that this measurement is well-suited to
exploring such capsomere conformational changes.

The permeation parameter is not irrelevant for making quantitative measurements. This parameter shifts the
value of the elastically dominated low-frequency plateau of elastic shells (such as viruses) by as much as
a factor of four. This dynamical response function is principally controlled by the bending modulus of the shell
so that AFM-based measurements of that quantity can be in error by as much as a factor of ten if permeation is neglected
in the interpretation of the data.  The effect of solvent permeation is generically more important in systems that
are both larger and mechanically more compliant. Thus the effect of solvent permeation can play a large role in the
mechanical response of large GUVs containing pore-forming proteins.

The combined effects of solvent permeation and viscoelasticity remains to be fully explored in a wide variety of
biological and synthetic nanoscale shells and vesicles. Based on our present work, accounting for solvent permeation
will be important for the quantitative interpretation of such rheological data. By taking such effects into account it
may be possible to make sensitive measurements of the collective effects of protein conformational change in viral
capsids and to probe the mechanics of a number of other similar structures.

\section*{Acknowledgements}
TK and AJL thank R.Bruinsma, W.Gelbart,
W. Klug, C.Knobler, and J. Rudnick for enjoyable and enlightening discussions.


\appendix

\section{The permeation coefficient}
\label{Appendix-A}
To determine the permeation coefficient we treat the porous membrane
of thickness $t$  as having an array of pores or radius $r_0$. We use
the well-known result for the fluid flux $Q$ through such a tube
of length $t$ and radius $r_0$ due to a pressure difference  $\Delta
P$ across the tube to write~\cite{Lamb:32}
\begin{equation}
Q=\frac{\pi r_0^4}{8\eta} \frac{\Delta P}{t}.
\end{equation}
We then suppose that the number of such tubes is $k$ per unit area
of the membrane and assume that the total flux through these
tubes is simply additive. Then the total flux through the area $L^2$
is thus given by
\begin{equation}
\mbox{total flux}= k L^2 Q,
\end{equation}
and the average velocity of fluid coming out of the membrane is
\begin{equation}
\label{v-average-app}
 v_{av}= k \frac{\pi r_0^4}{8\eta} \frac{\Delta
P}{t}.
\end{equation}
On the other hand, we assert from Darcy's law that
\begin{equation}
\label{Darcy-app}
v_{av}=\frac{\Delta P}{\alpha}.
\end{equation}
From Eqs.~(\ref{v-average-app}) and (\ref{Darcy-app}) we find the
permeation coefficient to be given by
\begin{equation}
\alpha=\frac{8\eta t}{\pi r_0^4 k}.
\end{equation}
Expected values of the permeation coefficient for a variety of
micron and nanoscale hollow, porous shells are given in Table 1. In
the text we introduce a dimensionless form of the permeation
coefficient $\bar{\alpha}$ that is defined by
\begin{equation}
\label{bar-alpha}
\bar{\alpha}\equiv \frac{\alpha\cdot R}{\eta} = \frac{8\, R\, t}{\pi
r_0^4\, k},
\end{equation}
where $R$ is radius of the sphere and $\eta$ the viscosity of the
surrounding solvent.

\section{Undulations of a plane permeable membrane}
\label{Appendix-B}
We calculate the fluid flow associated with the undulations of a
permeable membrane. To begin we look for solutions of the
incompressible Stokes equation
\begin{eqnarray}
\label{incompress-app}
&\Nabla \cdot{\bf v}=0 \\
\nonumber &\displaystyle \rho \frac{\partial {\bf v}}{\partial t}=\eta
\nabla^2 {\bf v} - \Nabla p. \label{stokes-app}
\end{eqnarray}
driven by sinusoidal membrane undulations of the form
\begin{equation}
\label{undulations-app}
 w=w_q e^{i q x - i \omega t}.
\end{equation}
These undulations apply an external normal stress on the fluid at
the membrane located in the $xy$ plane ($z=0$) of the form
\begin{equation}
\pi_e=\pi e^{i q x - i \omega t}.
\end{equation}
Correspondingly, we expect a solution for the fluid velocity field
${\bf v}$ and pressure $p$ of the form
\begin{eqnarray}
&{\bf v}={\bf v}(z)e^{i q x - i\omega t}\\
& p=p(z)e^{i q x - i\omega t},
\end{eqnarray}
but the fluid velocity and pressure fields must vanish far from the
membrane $z \to \pm \infty$ and satisfy at the membrane
\begin{eqnarray}
&v^+_x(0)=v^-_x(0)=0&\\
&v_z^+(0)= v_z^-(0)&
\end{eqnarray}
Additionally, the fluid incompressibility condition gives
\begin{equation}
\label{incompr} \frac{\partial v_z}{\partial z} + v_x (z) i q =0.
\end{equation}
Using these in Eq.~(\ref{stokes-app}) equation we find
\begin{eqnarray}
&& -\rho \,i \omega \,v_x(z)=\eta \left(-q^2 v_x +
\frac{\partial^2 v_x}{\partial z^2 }\right)- i\, q \,p(z)\label{NS1}\\
&& -\rho \, i \omega\, v_z(z)=\eta
\left(-q^2 v_z + \frac{\partial^2 v_z}{\partial z^2 }
\right)-\frac{\partial p(z)}{\partial z}\label{NS2}
\end{eqnarray}
The general form of the solutions to these ordinary differential
equations is
\begin{equation}
v_x(z)=a e^{\lambda z} \quad v_z(z)=b e^{\lambda z} \quad p(z)=c
e^{\lambda z},
\end{equation}
where the incompressibility condition Eqs.~(\ref{incompress-app}) and
(\ref{incompr}) requires that the undetermined constants are
related by
\begin{equation}
b=-\frac{i q}{\lambda} a,
\end{equation} while Eqs~(\ref{NS1}) and (\ref{NS2}) give four solutions
\begin{equation}
\lambda^2 = q^2,\, q^2 - i \frac{\rho \omega}{\eta}.
\end{equation}
Using the boundary conditions we find solutions of the form
\begin{eqnarray}
\label{vx-sol-app}
&v_x^\pm(z)&= \mp i q A (e^{\mp q z} -e^{\mp lz}) e^{i q x-i\omega t}\\
&v_z^\pm(z)&=q A (e^{\mp q z} -\frac{q}{l} e^{\mp l z}) e^{i q x-i\omega t}\\
&p^\pm(z)&= \mp i \omega \rho A e^{\mp q z} e^{i q x-i\omega t},
\label{vz-sol-app}
\end{eqnarray}
with $l=\sqrt{q^2-i\rho\omega/\eta}$.

Knowing the pressure and velocity fields, we require force balance
at the membrane:
\begin{equation}
(\sigma_{zz}^- -\sigma_{zz}^+) + \pi =\kappa \nabla^4 w,
\end{equation}
using the fluid stress tensors and membrane bending modulus
discussed in the text.

The permeability of the membrane sets the velocity difference
between the fluid and the membrane at $z=0$. Thus we also have
\begin{equation}
\alpha \left ( v_z(0)- \frac{d w}{dt} \right )=(p^- -p^+).
\end{equation}
These conditions set the two remaining unknown constants above, $A$
from Eqs.~(\ref{vx-sol-app})-(\ref{vz-sol-app}) and $w_q$ by requiring nontrivial solutions to
the set of algebraic equations
\begin{eqnarray}
&& A\left[ \alpha q \left(1-\frac{q}{l} \right ) -2\,i\omega \rho \right] +
w_q\,\alpha \, i \omega =0\nonumber \\
&& A\,2i\omega\rho - \kappa q^4 w_q=-\pi.
\label{sys_algebraic1}
\end{eqnarray}
Working in the limit where fluid inertia is irrelevant $\omega\rho\ll \eta q^2$ and solving for
$w_q$ we find
\begin{equation}
\displaystyle w_q=-\frac{ \pi\, \left[ \frac{\alpha
}{2q\eta}+2\right]} {2i\omega\alpha -\kappa q^4 \, \left[
\frac{\alpha }{2q\eta}+2 \right]}
\end{equation}
as discussed in the text.

\section{The A matrix}
\label{A-mat-app}
The eigenvectors and eigenvalues of the $A$ matrix
defined by Eq.~(\ref{eq:w-psi}) determine the relaxation
times and mode structure of the coupled membrane deformation and
solvent flow modes. The components of this matrix are listed below:

\begin{eqnarray}
\lefteqn{A_{11}=(2\ell+1)(4\ell^3+6\ell^2-\ell+3)\bar{\alpha}}\\
&&{}-(\ell -1)^2(\ell+2)^2(2\ell+1)(4\ell^3+6\ell^2-\ell+3)
\frac{1}{i\,\bar{\omega}_{\ell}} \nonumber \\
&&{}-(\ell-1)^2\ell(\ell+1)(\ell+2)^2(2\ell+1)\frac{\bar{\alpha}}{i\,\bar{\omega}_{\ell}}
\nonumber \\
&&{}-4(2\ell+1)(4\ell^3+6\ell^2-\ell+3)(2\ell+1)\frac{\bar{K}}{i\,\bar{\omega}_\ell}\nonumber \\
&&{}-2\ell(\ell+1)(\ell+2)\frac{\bar{\alpha}\,\bar{K}}{i\,\bar{\omega}_\ell}, \nonumber \\
\lefteqn{A_{12}=\ell(\ell+1)\Bigl[-3(2\ell+1)\bar{\alpha} } \\
&&{}+2\,(2\ell+1)(4\ell^3+6\ell^2-\ell+3)\frac{\bar{K}}{i\,\bar{\omega}_\ell} \nonumber \\
&&{}-3(\ell-1)\ell(\ell+2)\frac{\bar{\alpha}\,\bar{\mu}}{i\,\bar{\omega}_{\ell}}\Bigr], \nonumber
\end{eqnarray}
and
\begin{eqnarray}
\lefteqn{A_{21}=-3(2\ell+1)\bar{\alpha}} \\
&&{}+2(\ell-1)^2(\ell+2)^2(2\ell+1)\,\frac{1}{i\,\bar\omega_{\ell}}\nonumber\\
&&{}-3(\ell-1)^2\ell(\ell+2)^2\,\frac{\bar{\alpha}}{i\,\bar{\omega}_{\ell}}\nonumber\\
&&{}+12(2\ell+1)\,\frac{\bar{K}}{i\,\bar{\omega}_{\ell}}\nonumber\\
&&{}+2\ell(\ell+2)(2\ell-1)\frac{\bar{\alpha}\bar{K}}{i\,\bar{\omega}_{\ell}}\nonumber\\
\lefteqn{A_{22}= \ell(2\ell+1)(4\ell^2+6\ell-1)\bar{\alpha}}\\
 &&{}-6\ell(\ell+1)(2\ell+1)\frac{\bar{K}}{i\,\bar{\omega}_\ell} \nonumber \\
&&{}-\ell^2(\ell+1)(\ell+2)(2\ell-1)\,\frac{\bar{\alpha}\bar{K}}{i\,\bar\omega_{\ell}}\nonumber\\
&&{}-(\ell-1)\ell(\ell+2)(2\ell^2+3\ell+4)\,\frac{\bar{\alpha}\bar{\mu}}{i\,\bar\omega_{\ell}}
\nonumber.
\end{eqnarray}

\section{The response function}
\label{response-app}

Here we list the $\ell$ dependence of each function entering into
the diameter response function shown in Eq.(\ref{response_function}). These
functions depend also on the dimensionless permeation parameter
$\bar{\alpha}$, dimensionless shear modulus $\bar{\mu}$ and dimensionless area compression modulus $\bar{K}$.

\begin{eqnarray}
\lefteqn{\Lambda_1(\ell )=2(1+2\ell)(6+7\ell+7\ell^2)\bar{K} }  \\
&&{}+\ell (\ell +1)(\ell^2+\ell +4)\,\bar{\alpha}\,\bar{K} \nonumber \\
&&{}+(\ell-1)\ell (1+\ell)(2+\ell)\bar{\alpha}\bar{\mu} \nonumber \\
\lefteqn{\Lambda_2(\ell )=2(1+2\ell )(3+\ell +\ell^ 2)\bar\alpha}\\
\lefteqn{\Lambda_3(\ell )=(\ell-1)^2\ell(1+\ell)(2+\ell)^2\times}\\
&&(1+2\ell)(3+2\ell+2\ell^2)\bar{K}\nonumber\\
&&{}+(\ell-1)^2\ell^2(1+\ell)^2(2+\ell)^2\bar{\alpha}\bar{K}\nonumber\\
&&{}+(\ell-1)^3(2+\ell)^3(1+2\ell)(3+2\ell+2\ell^2)\bar{\mu}\nonumber\\
&&{}+(\ell-1)^3\ell(1+\ell)(2+\ell)^3\bar{\alpha}\bar{\mu}\nonumber\\
&&{}+4(\ell-1)(2+\ell)(1+2\ell)(3+2\ell+2\ell^2)\bar{K}\bar{\mu}\nonumber\\
&&{}+4(\ell-1)\ell(1+\ell)(2+\ell)\bar{\alpha}\bar{K}\bar{\mu}\nonumber\\
\lefteqn{\Lambda_4(\ell )=-(2\ell +1)\Bigl[(\ell -1)^2(\ell +2)^2\times}\\
&&{}(2\ell -1)(2\ell +1)(2\ell +3) \nonumber \\
&&{}+2(\ell -1)^2 \ell  (\ell +1)(\ell +2)^2\,\bar{\alpha} \nonumber \\
&&{}+4(2\ell -1)(2\ell +1)(2\ell +3)\,\bar{K}\nonumber \\
&&{}+\ell(1+\ell)(2\ell^2+2\ell-1)\,\bar{\alpha}\bar{K}  \nonumber \\
&&{}+(\ell-1)(2+\ell)(3+2\ell+2\ell^2)\bar{\alpha}\bar{\mu}\Bigr ] \nonumber\\
\lefteqn{\Lambda_5(\ell )=-(2\ell -1)(2\ell +1)^2(2\ell +3)\bar{\alpha}}
\end{eqnarray}


\end{document}